\documentclass[12pt,preprint,letter]{aastex}

\newcommand{\trans}[2]{J={#1}-{#2}}

\slugcomment{}

\shorttitle{Remnant Outflow in G5.89}
\shortauthors{Klaassen et al.}

\begin{document}

\title{The Possibly Remnant Massive Outflow in G5.89-0.39 : I - Observations and Initial MHD Simulations}

\author{P. D. Klaassen\altaffilmark{1}, R. Plume, R. Ouyed, A. M. von Benda-Beckmann\altaffilmark{2}}
\affil{Dept. of Physics and Astronomy, University of Calgary, Calgary, AB, Canada}
\email{pamela@ism.ucalgary.ca}

\and

\author{J. Di Francesco\altaffilmark{3}}
\affil{National Research Council Canada, Herzberg Institute of Astrophysics, Victoria, BC, Canada}

\altaffiltext{1}{Currently at McMaster University, Hamilton, ON, Canada}
\altaffiltext{2}{Currently at Astrophysical Institute Potsdam, Potsdam, Germany}
\altaffiltext{3}{Also at University of Calgary}

\begin{abstract}
We have obtained maps of the large scale outflow associated with the UCHII region G5.89-0.39 in CO and $^{13}$CO (\trans{3}{2}), SiO (\trans{8}{7},\trans{5}{4}), SO$_2$ (\trans{13$_{2,12}$}{13$_{1,13}$}) and H$^{13}$CO$^+$(\trans{4}{3}).  From these maps we have been able to determine the mass (3.3 M$_{\odot}$), momentum (96 M$_{\odot}$ km s$^{-1}$), energy ($3.5\times 10^{46}$ erg), mechanical luminosity (141 L$_{\odot}$), and mass loss rate ($\sim 1\times 10^{-3}$ M$_{\odot}$ yr$^{-1}$) in the large scale outflow. The observationally derived parameters were used to guide 3D magnetohydrodynamic models of the jet entrained outflow.  Through the combination of observations and simulations, we suggest that the large scale outflow may be inclined by approximately 45$^{\circ}$ to the line of sight, and that the jet entraining the observed molecular outflow may have been active for as little as 1000 years, half the kinematic age of the outflow.
\end{abstract}


\section{Introduction}
\label{sec:intro}

Over the past 20 years there have been significant advances in the study of star formation. Several phases of the star formation process have been identified and characterized, beginning with a centrally concentrated core of molecular gas, which collapses to form a star surrounded by a proto-planetary disk (see for example, Shu et al. 1987).  A large part of this progress has focused on the formation of low-mass stars, since they can form in isolation, are closer, and more abundant, reducing source confusion. Similarly, the formation timescales are long enough ($t_{\rm{form}}\gtrsim5\times10^5$ yr, e.g. Hartmann 2000) for a considerable number of these objects to be detectable.

Several studies have shown that  low mass protostars have evolutionary phases both with and without outflows (Class 0/I and Class II/III, e.g. Lada 1987, Andr\'e et al. 1993) and that younger outflows appear to be more collimated and energetic (e.g. Bachiller \& Tafalla 1999). A significant amount of effort has also gone into characterizing high mass protostellar phases but an analogous sequence has yet to be found. Meuller et al. (2002), and Shirley et al. (2002) used the same methods to determine physical properties in high and low mass star forming regions respectively and found accretion rates for high mass  ($M > 8$ M$_{\odot}$) protostars were at least three orders of magnitude greater than those for low mass protostars. While in low mass systems, the Kelvin-Helmholtz timescale ($t_{\rm KH}=GM_*/R_*L_*$) is too long to be of interest, in high mass systems it can become shorter than the free fall timescale, possibly causing the protostar to begin to radiate before fully accreting all of its material (Cesaroni 2004). Because high mass protostars are further away on average than their low mass counterparts, and are deeply embedded objects, they are generally studied through their effects on their environments, e.g., their molecular outflows.  Studies of the outflow mechanisms can provide valuable insight into the accretion processes for low mass protostars since the two processes appear related (e.g. Andr\'e et al. 1993). This could be similarly true for high mass protostars but the process may be complicated by such things as the radiation pressure and ionizing UV flux from the central star. 

The larger average distances, shorter accretion times and the clustered nature of massive star formation (e.g. Lada, Bally \& Stark 1991) reduce our ability to isolate individual, massive, pre-protostellar cores.  As a result, we still do not understand how massive stars form. Is it simply a scaled-up version of low-mass star formation in which accretion rates are high enough to overcome radiation pressure (e.g. McKee \& Tan 2003; Cesaroni 2004), or do massive stars form from the coalescence of lower mass protostars (e.g. Bonnell et al. 2001)? To investigate the outflow phenomenon in massive star formation,  we will focus on the molecular outflow in G5.89-0.39 (hereafter G5.89) - also known as W28A2.

 The ultracompact HII (UCHII) region in G5.89 (cataloged by Wood \& Churchwell 1989) has a radius of 0.01 pc, is powered by an O5 ZAMS star (Feldt et al. 1999) and is expanding at a rate of $\sim35$ km s$^{-1}$ (at an assumed distance of 2 kpc, Acord et al. 1998).   The molecular outflow in G5.89 was first studied by Harvey \& Forveille (1988, hereafter HF88) who mapped the innermost square arcminute of the region and found that, at the edge of their map, high velocity line wings  ($\Delta v_{\rm{FWZP}}\approx$ 25 km s$^{-1}$ in $^{13}$CO) were still apparent, suggesting the molecular outflow was much more extended than the area covered by their map.  Recent sub-millimeter studies of this region have placed additional constraints on properties of the flow, such as momentum, kinetic energy, mechanical luminosity, age, mass loss rate, and force in both the ambient material and outflow lobes using the shock tracer SiO (Sollins et al. 2004; Acord et al. 1997, hereafter AWC97), showing this source to be quite powerful. The broad line wings, and extremely high mass powering source make this an interesting outflow to characterize. In terms of outflow energetics, the G5.89 outflow is the sixth most energetic in the Wu et al. (2004) study of high velocity molecular outflows, placing it in the top 2\% of their outflows with energy calculations.

In Section \ref{sec:Observations} we present the first observations of the full extent of the molecular (CO) outflow in G5.89, as well as SiO and serendipitous SO$_2$ and H$^{13}$CO$^+$ observations.  We use these data to derive a number of physical properties of the outflow,  which we use to constrain simulations of a jet entrained molecular outflow in Section \ref{sec:Simulations}. These high resolution magnetohydrodynamic (MHD) simulations test the number of protostellar sources required to power the observed outflow, the inclination angle of the outflow, and the jet lifetime required to achieve the observed dynamics. Comparisons between observations and simulations will be drawn in Section \ref{sec:discussion}. We then summarize our results in Section \ref{sec:conclusions}.  For all calculations in this paper, we have adopted the more recent distance estimate to G5.89 of 2 kpc (i.e. Acord et al. 1998, Feldt et al. 2003), noting that this will cause systematic differences in derived properties from previous studies.

\section{Observations and Results}
\label{sec:Observations}

Observations of the rotational transitions of $^{12}$CO \trans{3}{2}, $^{13}$CO \trans{3}{2}, SiO \trans{8}{7} and SiO \trans{5}{4} were taken in  2003 April and May at the James Clerk Maxwell Telescope (JCMT)\footnote{The JCMT is operated by the Joint Astronomy Center on behalf of the Particle Physics and Astronomy Research Council of the United Kingdom, the Netherlands Organization for Scientific Research, and the National Research Council of Canada.} on Mauna Kea, Hawaii.  CO is a tracer of the bulk of the gas in the interstellar medium, since it is the second most abundant molecule, and the most abundant molecule with dipole rotational transitions (e.g. Bachiller, 1996). We discuss our CO observations below in Section \ref{subsec:CO}. SiO is generally used to trace recently shocked gas, since silicon is generally frozen out of the gas phase in the ISM (e.g. Martin-Pintado et al. 1992).  The transitions of  SO$_2$ (\trans{13$_{2,12}$}{13$_{1,13}$}) and H$^{13}$CO$^+$ (\trans{4}{3}) were serendipitously observed during the SiO \trans{5}{4} and \trans{8}{7} observations respectively, as discussed in Section \ref{sec:SIO}.

The central position of each map was the location of the Cesaroni et al. (1988) water maser at W28A2 (1); $\alpha_{B1950}$ = 17$^h$57$^m$26$.^s$803, $\delta_{B1950}$ = -24$^{\circ}$03$'$54$.''$02.  Feldt et al. (2003) recently identified a protostar at $\alpha_{2000}=18^{\rm{h}}00^{\rm{m}}30.^{\rm{s}}44\pm0.^{\rm{s}}013$, $\delta_{2000}=-24^{\circ}04'0''.9\pm0''.2$ as the source of the UCHII region, but do not comment on whether this is the source of the large scale outflow. This is offset from the water maser by 2$''$.4, and does not correspond to any other water maser in the Cesaroni et al. list. Our map center lies south of the Feldt et al. (2003) center, in a region they show to be highly obscured by dust.  All observations were reduced using the SPECX (Prestage et al., 2000) and CLASS (Buisson et al. 2002) software packages.

Observations with Receiver A (230 GHz) were made in double sideband mode for SiO (\trans{5}{4}), had a half power beamwidth (HPBW) of 20$''$, and a main beam efficiency ($\eta_{mb}$) of 0.62. Observations made with Receiver B (345 GHz) were in single sideband mode for the $^{12}$CO, $^{13}$CO, and SiO \trans{8}{7} observations, had a HPBW of 14$''$, and $\eta_{mb}$ = 0.62. All observations were taken in raster mapping mode, with the DAS configured to 760 MHz. Observations were coadded, first order baselines were subtracted, and the lines were fit with Gaussian line profiles.  Table \ref{tab:obs_param} provides a summary of our JCMT observations of the G5.89 region.

\subsection{$^{12}$CO and $^{13}$CO}
\label{subsec:CO}

We removed the Gaussian fit line centers from our spectra, leaving only the residual outflow component of the emission, the so-called {\it line wing} emission.  Unless stated otherwise, when describing integrated intensities, we are referring to the integrated intensity of the residual line wing emission (either red shifted or blue shifted), not the emission from the line center.  Gaussian models were fit to each $^{12}$CO and $^{13}$CO spectrum individually, using the Gaussian model routine within CLASS, and removed using the RESIDUAL command as shown for $^{13}$CO at our map center in Figure \ref{fig:residual}.  For reference, the average $^{12}$CO Gaussian parameters with standard deviations were: $V_{LSR}$ = 8.3$\pm$1.0 km s$^{-1}$, full width half maximum (FWHM) = 4.6$\pm$1.4 km s$^{-1}$ and $T_R^*$ = 26.7$\pm$11.1 K.  The velocity extent of the wings was set as the range from -66 km s$^{-1}$ to 78 km s$^{-1}$, which is the full width at zero power towards the central position.  Using this residual method, we have been able to separate the cloud emission from the outflow emission. The $^{13}$CO emission is very well fit by a Gaussian with line wings, suggesting that this would also be the case for the $^{12}$CO if the self-absorption were not present.  The blue side of the $^{12}$CO emission does appear to be well fit with a Gaussian and line wing emission.  Removal of the Gaussian from the $^{12}$CO emission did tend to give negative intensities within the lower velocity red shifted absorption trough. To avoid incorporating these negative intensities into our subsequent calculations, we began calculating integrated intensities redward of 20 km s$^{-1}$. This velocity was chosen because it is well outside of the 1 $\sigma$ error bars on the average FWHM of the Gaussian, yet incorporates as much red shifted emission as possible.   The presence of the absorption features will, unfortunately, tend to underestimate the integrated intensity of the line wing.

To constrain better the properties of the G5.89 outflow over previous maps, our fully sampled $^{12}$CO (\trans{3}{2}) map extends $3'$ along the flow axis, and $2'$ perpendicular to it, while our $^{13}$CO  (\trans{3}{2}) map covers the innermost $98''\times98''$ of the region. Figure \ref{fig:12COmap} shows the total integrated intensity and the residual line wing emission of the $^{12}$CO and $^{13}$CO \trans{3}{2} towards G5.89.  We define the edges of the outflow to be the locations in the CO map where the line wing integrated intensity is 10\% of its maximum value. The maximum values are $\int T dv_b$ = 642 K  km s$^{-1}$, $\int T dv_r$ = 452 K km s$^{-1}$ for the blue and red wings respectively, while the rms noise is 6 K km s$^{-1}$.  From this definition, the $^{12}$CO outflow subtends just under 2$'$ (49$''$ along the red lobe, and 56$''$ along the blue) along the flow axis, and just under 1$'$ perpendicular to it and our maps encompass the entire outflow. In the less optically thick tracer ($^{13}$CO), the full extent of the outflow is 80$''$ along the flow axis, and 50$''$ perpendicular to it. At a distance of 2 kpc, the $^{12}$CO outflow appears to extend 1.2 pc on the sky (without consideration for inclination angle effects).  A literature search revealed no constraints on the inclination angle of this outflow.  The outflow does however appear extended on the sky with little overlap between the red and blue lobes, thus it is likely that the outflow is not primarily oriented along the line of sight (i.e. 90$^{\circ}$ inclination). We compare our observations to the models of Cabrit \& Bertout (1990) in Section \ref{sec:observed_properties}.

Every position in the $^{12}$CO map contains absorption due to cold clouds along the line of sight at $V_{\rm{LSR}}$=13.7 km s$^{-1}$ and 20.2 km s$^{-1}$. The presence of these absorption features affects the total observed integrated emission from the red shifted outflow which produces uncertainties in calculations requiring integrated intensities.  Figure \ref{fig:spectra} shows spectra from the central position in each tracer and the two line of sight absorption features in $^{12}$CO are clearly distinguishable. Figure 1 of HF88 shows three CO isotopes (CO, $^{13}$CO, and C$^{18}$O in the \trans{1}{0} transition) and two CS isotopes (CS, and C$^{34}$S in the \trans{2}{1} transition). In both HF88 and our study, $^{12}$CO is the only molecular species which appears to suffer from the two absorption features.  Thus we assume our $^{13}$CO observations are free of line-of-sight absorption.  Note, however, that while Figure \ref{fig:12COmap} shows a red outflow lobe which is less extended than the blue lobe, this is not likely to be an effect of the absorption features. Since the red absorption features appear at every position in the map, calculating the extent of the outflow as a percentage of the maximum causes the absorption effect to cancel out.  The same is {\it not} true of calculations of physical quantities, however (see Section \ref{sec:observed_properties}). 

Figure \ref{fig:w28a2PV} shows the position-velocity (PV) diagram of $^{12}$CO  emission along the flow axis ($\Delta\delta$ = 0).  It reveals that the highest velocity gas is concentrated towards the center of the outflow, suggesting that the gas decelerates away from the source (see for instance Cabrit \& Bertout 1986, 1990). The solid lines show a $v\propto r^{-1}$ trend, which the bulk of the gas tends to follow.  For reference, the dashed line shows a $v\propto r$ trend, which represents accelerating gas.  At very low levels on the blue shifted side of the PV diagram, there does appear to be some accelerating gas in this region.  We suggest that this may be a contribution from the accelerating outflow in this region detected by Sollins et al. (2004). Their outflow (in SiO) appears to have a velocity extent of $\approx$ 50 km s$^{-1}$ and an angular extent of 15$''$.  Our tenuously detected second (accelerating) outflow appears to extend 60$''$ in CO, beyond their primary beam.  However, since we do not have the spatial resolution to confirm this hypothesis, we do not discuss it further.

\subsection{SiO, SO$_2$ and H$^{13}$CO$^+$}
\label{sec:SIO}

Silicon is a key constituent of dust grains, and the passage of a shock wave can remove Si from dust grains via sputtering (e.g. see Field et al. 1997).  The elevated temperatures arising from a shock can also liberate Si-bearing species from frozen grain mantles.  Once in the gas phase, Si atoms react quickly with oxygen, forming SiO.  SiO, however, can also be quickly oxidized into SiO$_2$, or freeze back onto grain mantles in $< 10^4$ years (e.g. Pineau des Forets et al. 1997).  Thus, SiO is an excellent tracer of recent shocks, such as those produced by molecular outflows (e.g. Bachiller 1996). The combination of CO and SiO observations, therefore, can yield a broader understanding of the G5.89 outflow.

Figure \ref{fig:total_intensity} shows $80''\times80''$ maps of the SiO \trans{5}{4} and {\trans{8}{7}. In the SiO \trans{5}{4} data we suggest that the  serendipitous line seen at $V_{LSR} \sim  77$ km s$^{-1}$ is SO$_2$ \trans{13$_{2,12}$}{13$_{1,13}$} in the upper sideband ($\nu=225.154$ GHz), as seen in a 215-247 GHz line survey of Orion A (Sutton et al. 1985). Models of shock propagation and chemical evolution concerning sulfur bearing molecules show that at temperatures in the 100-300 K range, SO$_2$ is the dominant sulfur bearing species   (Doty et al. 2002, Charnley, 1997). The SO$_2$ emission in our observations appears to have the same spatial extent as the SiO emission, suggesting a common origin (shocked, warm gas). Another serendipitous line was detected in the SiO \trans{8}{7} observations, and is shown in the inset of Figure \ref{fig:spectra}. This line is H$^{13}$CO$^+$ \trans{4}{3} at $\nu=346.999$ GHz, consistent with the molecular line survey of G5.89 by  Thompson \& MacDonald (1999).

\subsection{Physical Parameters Derived from Observations}
\label{sec:observed_properties}

Assuming that the $^{12}$CO emission is optically thick, that the level populations are in LTE, and a beam filling factor of one, we derive a kinetic temperature at the central position of $\sim$ 80 K from the peak brightness of the line.  We note however, that one transition of CO is not a sensitive temperature tracer (especially of material deeper into the cloud than the $\tau \sim 1$ surface) and suggest that this temperature is a crude lower limit to the actual kinetic temperature of the region. Observations of NH$_3$ were used by AWC97 to place a lower limit on the temperature of the SiO emitting region at 100 K.  The presence of SO$_2$ in our observations also suggests a lower limit of $T_{K}\sim100$ K (Doty et al. 2002).  Thus, we have adopted a kinetic temperature of 100 K for the central position of G5.89.

Schilke et al. (1997) presented a set of C-type chemical shock models with a variety of different ambient densities and shock velocities. Their results include level populations up to SiO J$_{\rm up} =$ 15 (see their Figure 6) as well as integrated intensities and line ratios for three of the SiO rotational transitions observable by ground based telescopes (J=2-1, 5-4, and 8-7). Convolving our \trans{8}{7} observations to the beamsize of our \trans{5}{4} observations, we find a line  integrated intensity ratio of [8-7]/[5-4] = 0.83. Comparing this to the Schilke et al.  model, we estimate $n_{\rm{H}}=10^7$ cm$^{-3}$ and $v_s$=28 km s$^{-1}$, but note that at $v_s\gtrsim 30$ km s$^{-1}$, this ratio is only weakly dependent on shock speed. A simple large velocity gradient (LVG) model of the SiO emission from the two observed transitions, however, gives similar ambient densities ($n=1.4\times10^7$ cm$^{-3}$). These densities are higher than those derived by previous studies ($n\approx 10^6$ cm$^{-3}$; Plume et al. 1997, AWC97). 

The observed outflow lobes in G5.89 are not highly inclined towards the line of sight, and a comparison of Figures \ref{fig:12COmap} and \ref{fig:w28a2PV} to the models of decelerating outflows of Cabrit \& Bertout (1990), show that this outflow lies somewhere between their $i=10^{\circ}$ and $i=50^{\circ}$ cases, closer to the $i=50^{\circ}$ case.  We suggest an inclination angle of approximately 45$^{\circ}$ with respect to the plane of the sky.

The dynamical properties of the outflow (such as momentum, energy, luminosity, and age) are best studied when the line emission is broken down into velocity intervals, allowing for examination of the mass and dynamics in each velocity bin.  Summing over all the bins gives the total momentum, energy, etc., in each outflow lobe. Figure \ref{fig:channel} shows channel maps of the $^{12}$CO emission from G5.89, while Table \ref{tab:results} shows the dynamical properties calculated, the equations used, and values obtained for each outflow lobe. The kinematic age of the outflow was determined by dividing the extent of the outflow (scaled by a factor of $\sin 45^{\circ}$ for the inclination angle) by the average gas velocity (scaled by a factor of $\cos 45^{\circ}$ for the inclination angle), resulting in an age of $t_{\rm kin}$ = $t_{\rm kin}^{\rm proj}\tan 45^{\circ}\sim$ 2000 yr. This age is consistent with that given by Wu et al. (2004), who used similar methods to derive outflow lifetimes.

Using the column density estimated from  the $^{13}$CO observations, an assumed $^{12}$CO/$^{13}$CO abundance ratio of 55 (Langer et al. 1984), and the derived abundance of H$_2$ in this region with respect to CO (1.7$\times 10^4$; van der Tak et al. 2000), we can determine the abundances of our other molecular species.  Based on our SiO \trans{8}{7} observations, which have a higher signal to noise ratio than our \trans{5}{4} observations, and the same LTE approximation used to determine the $^{13}$CO column density (such as an ambient temperature of 100 K), we find [SiO]/[H$_2$] = 3$\times 10^{-10}$. This abundance is consistent with SiO being recently released from grain mantles into the gas phase (Schilke et al. 1997). Using the RADEX online calculator\footnote{http://www.strw.leidenuniv.nl/$\sim$moldata/radex.php}, we were able to determine the column density of our SO$_2$ gas (N$_{\rm SO_2}\sim 3\times 10^{15}$ cm$^{-2}$), and an abundance ratio of [SO$_2$]/[H$_2$] = 2$\times 10^{-8}$.  We compare this SO$_2$ abundance to the hot core sulfur models of Charnley (1997), to find that SO$_2$ reaches this abundance after approximately 1900-2500 yr. This age is consistent with the kinematic age of the outflow described above.

 If we assume an accretion rate 10 times the (constant) outflow rate (shown in Table \ref{tab:results}), it would take 2000 years for 60$M_{\odot}$ (the mass of an O5 star) to accumulate.  Studies of accretion and outflow rate ratios suggest that even higher ratios exist (e.g. Beuther et al. 2002), and thus it is possible for the accretion timescale to be shorter than 2000 yr.

We can easily compare our results to those of Beuther et al. (2002).  They determined physical parameters for 33 massive molecular outflows using methods similar to ours.  They did use a slightly higher $^{13}$CO abundance (8.9$\times10^5$) than we did (3.2$\times10^5$), but when we correct for this, we find that our mass, momentum and mass loss rates are lower than their average values, but that our energy, luminosity and force are all significantly higher than their average values.  There are only two sources in their source list (18264-1152 and 19410+2336) which have higher outflow energies than G5.89, and only one (19410+2336) which has a higher mechanical luminosity and force.  Our derived values for this source are consistently smaller than those derived by previous studies (i.e. HF88, AWC97, Sollins et al. 2004).  The authors of those studies each used the total integrated intensity of their observed lines, whereas we only used  the outflowing line wing emission to determine outflow properties.  

To further the analysis of the G5.89 flow, we used the observationally determined properties of the outflow to guide magnetohydrodynamic (MHD) simulations of a disk-wind driven molecular jet which entrains the ambient molecular gas into an outflow.  This allows us to produce an outflow mimicking our observations, from which we can draw more information, and conclusions about the nature of massive star formation. 

\section{Simulations and Results}
\label{sec:Simulations}

There are a number of theories suggesting entrainment mechanisms of molecular outflows associated with massive star formation. While we do not discuss the specifics of the driving mechanism, we assume an accretion model, and that a disk wind driven molecular jet (e.g. Pudritz \& Norman, 1986, Raga \& Cabrit 1993, Masson \& Chernin 1993) entrains the surrounding molecular gas. Regardless of the specifics of jet generation and outflow entrainment, we can estimate the inclination angle, kinematic age, and length of time the jet is powering the outflow by comparing our MHD simulations to observations.  We can use observational constraints as initial conditions for the simulations.

There are two distinct sets of input parameters required for our simulations: source and ambient.  The source parameters were based on an O5 zero age main sequence (ZAMS) star, like the one powering the G5.89 outflow (Feldt et al. 1999).  Ambient parameters were fixed based on the observational analysis presented above.  All of these parameters are given in Table \ref{tab:init_sim}. Parameters derived from these quantities are discussed below.

Within our model, the outflow particles are attached to the magnetic field lines, and so, are not launched from the surface of the protostar, but from the magnetic footpoints in the disk ($r_o$).  They are then accelerated until they reach the Alfv\'enic radius ($r_a$) where collimation of the outflow occurs. It is the ratio between these two radii ($r_o/r_a$) which determines the amount of collimation and acceleration within the jet (e.g. Blandford \& Payne 1982, Pudritz \& Norman 1986).  We will define this ratio to be $\alpha^2$, resulting in $\alpha$ becoming the scaling factor between the Keplerian velocity at the edge of the star, and the point at which the particles are launched into the outflow (Ouyed \& Pudritz, 1997):

\begin{equation}
v_w = \alpha \sqrt{\frac{2GM_*}{R_*}}
\label{eqn:wind_vel}
\end{equation}

\noindent Setting $\alpha$ = 0.4 gives $v_w$ = 552 km s$^{-1}$.  This velocity is only slightly slower than the H$\alpha$ line wings in HH 444 discussed in Andrews et al. (2004). As a control, we also set $\alpha$ = 0.2, to see the effects of lower wind velocities on the final extent of the outflow.

In general, how much of the source bolometric luminosity is transfered into the outflow mechanical luminosity is a poorly constrained parameter. There are a number of ways of dissipating the energy produced by a protostar, such as momentum transfer to the outflow, shock heating of the ambient gas and atomic and molecular line cooling. Since the source is likely an O5 ZAMS star, we can constrain the bolometric luminosity to be $L_{\rm bol}=8\times10^5$ L$_{\odot}$ (Carroll \& Ostlie, 1996).  Similarly, from our outflow observations, we can constrain the mechanical luminosity of the flow to be $L_{\rm mech}=140$ L$_{\odot}$ (see Table \ref{tab:results}), and can express it as a fraction of the source bolometric luminosity ($L_{\rm mech}=\beta L_{\rm bol}$).  Other authors have suggested the possibility of multiple sources powering the outflow in G5.89 (i.e. Feldt et al. 2003, Sollins et al. 2004), however with our $\beta\approx10^{-4}$ we find, as do they, that we do not require a second source to account for the energetics of the outflow.

We observationally constrained (see Section \ref{sec:observed_properties}) the ambient density within the central 14$''$ of the G5.89 outflow to be $\sim 10^7$ cm$^{-3}$, and applied a Plummer profile  covering a region of radius 0.6 pc to mimic density gradients common to protostellar regions, which results in an average density of 10$^4$ cm$^{-3}$ (i.e. Whitworth \& Ward-Thompson 2001, Boily \& Kroupa 2003).  To simplify our initial simulations, we used the averaged density (10$^4$ cm$^{-3}$) for the entire region, with ambient density gradients being left to future study.  A summary of all of our observationally constrained initial conditions is shown in Table \ref{tab:init_sim}.

Further simplifying assumptions include use of only one molecular line coolant, CO, and that it is not destroyed by the jet bow shock.  CO is a coolant inherent to the code, and at the temperatures suggested by observations, is one of the strongest coolants in the cloud (e.g Smith \& Rosen 2003). Simplyfing the cooling may slightly inflate the CO line profiles, but not significantly.  We also do not treat the heating, destruction and reformation of CO by the jet as we are more interested in the extent and lifetime of the jet and surrounding outflow, than the chemistry produced by the jet.  The momentum transfered to the ambient medium is the quantity we are attempting to measure, which is independent of the nature of the medium (e.g. whether it is atomic or molecular).  Neither of these simiplifying assumptions should adversely effect our results.

\subsection{Computational Details}
\label{sec:comp_details}

Simulations were conducted using the Zeus-MP astrophysical fluid dynamic code (Norman 2000) on the CAPCA\footnote{Animations of these simulations (like the panels shown in Figure \ref{fig:t_off=1000}) can be viewed by following the ``Animations'' link at www.capca.ucalgary.ca.} computer cluster.  The cluster consists of 64, 2.4 GHz, Linux based processors connected by a 1 GB network.  Results of these simulations were imaged and analyzed using JETGET (Staff et al. 2004), an interactive data language (IDL) based visualization program for use with hierarchical data format (HDF) files produced by Zeus like codes in two or three dimensions.  

The simulations were run in three Cartesian coordinates, with the protostar comprising the central pixel (or grid zone) of the simulated region.  The grid used for these simulations was $210\times 240 \times 210$ pixels, with a resolution of $x_1 \times x_2 \times x_3 = (1.1 \times 5 \times 1.1) \times 10^{-3}$ pc, giving a total extent of $0.24\times 1.2\times 0.24$ pc in the plane of the sky.   For reference, the $x_1$ plane is perpendicular to the outflow axis, and corresponds to the line-of-sight in the observations. The $x_2$ plane is parallel to the outflow axis and corresponds to right ascension (in projection), while the $x_3$ plane  is perpendicular to the outflow axis, corresponding to declination  (in projection) for the large scale G5.89 outflow.

\subsection{Simulation Results}
\label{sec:sim_results}

To determine how long the jet actively powers the outflow, we ran a number of simulations with varying jet activity timescales.  For each simulation, the jet was turned off at $t_{\rm off}$, and the molecular outflow was allowed to evolve to the kinematic age of the system ($t_{\rm kin}$ = 2000 yr; Section \ref{sec:SIO}).  Simulations were run using $t_{\rm off}$ = 300 yr, 500 yr, 1000 yr, and $\infty$ (jet does not shut off within $t_{\rm kin}$). We then compared the size of the simulated outflow lobes at $t_{\rm kin}$ = 2000 yr to the size of the observed CO lobes (1.2 pc at a distance of 2 kpc uncorrected for inclination angle; Section \ref{subsec:CO}).  The shorter jet lifetimes ($t_{\rm off}$ = 300 and 500 yr) were used to test the simulations and ensure we were not getting jet activity timescales which were unreasonably short.  Given the ambient density of  10$^4$ cm$^{-3}$, the outflows powered by short lived jets ($t_{\rm off}$ = 300 and 500 yr) reach full extents of 0.7 and 0.9 pc, respectively. As expected, the spatial extents, and gas velocities of these two sets of simulations were too low when compared to the observed outflow in G5.89.

We found that setting $\alpha$ = 0.2 did not produce results consistent with the G5.89 outflow.  With $\alpha$ = 0.2, an outflow with a jet lifetime of $t_{\rm off} = \infty$ could not reach the extent of the observed outflow.  Thus, $\alpha$ was set at 0.4 for the rest of our simulations. We acknowledge the possibility that $\alpha \neq$ 0.4.  However, for $\alpha<0.4$, the simulated outflow will not reach the observed spatial extent of the outflow.  Higher wind velocities ($\alpha>0.4$) would decrease the jet activity timescale. Observational evidence for higher velocity winds from protostars is scarce.

Figure \ref{fig:t_off=1000} shows the results of simulations with longer $t_{\rm off}$, at various epochs in the outflow evolution. The $x_3$ plane (y axis) has been expanded (with respect to the $x_2$ plane, or x axis) in order to show the details of the jet. The first five panels show the evolution of the $t_{\rm off}$ = 1000 yr case in 400 yr steps, while the sixth panel (shown with a different color scheme) shows the final result for the $t_{\rm off} = \infty$ case. Both sets of simulations appear to reach the same spatial extent as the observed outflow. For the longer $t_{\rm off}$ cases, it becomes harder to differentiate between simulation sets. In both cases, the outflow appears to have the same extent, with the highest velocity gas varying by less than 1\% between simulations.  The collimation factor of the outflow in both cases is $\ell_{\rm length}$/$\ell_{\rm width}\sim 3$, consistent with other high mass protostellar outflows (see for example, Beuther et al. 2002).  The highest velocity outflow gas in both cases is $\sim$ 70 km s$^{-1}$, with the jet velocity peaking at $\sim 430$ km s$^{-1}$.  A noticeable difference between the simulations is the extent of the underlying jet.  In the $t_{\rm off}$ = 1000 yr case, the latest time panel of Figure \ref{fig:t_off=1000} (bottom, middle) only shows the jet towards the ends of the outflow.  In the $t_{\rm off} = \infty$ case, the jet extends from the central source to the edges of the outflow.

Figure \ref{fig:convol} shows the density of the simulations both at the resolution of the simulations, and convolved to the resolution of the JCMT beam at 345 GHz (simulated pixels convolved with a 15$''$ Gaussian). Unlike Figure \ref{fig:t_off=1000}, where the y axis was stretched to show the details of the jet, the x and y axes in this figure have the same scaling. The outflow lobes appear asymmetrical for two reasons: First, they represent  single channel maps centered at 4 km s$^{-1}$ (with a width of 4 km s$^{-1}$) with respect to the source velocity.  This means that the emission in the red outflow lobe dominates the emission in the blue outflow lobe. Second, we have introduced and inclination angle of 45$^{\circ}$ as described in Section \ref{sec:Observations}. This inclination angle appears to match well with observations.  It is clear from Figure \ref{fig:convol} that higher resolution images are required in order to resolve the jet, both spatially and kinematically.

Figure \ref{fig:dens} shows the density of the jet at the same times shown in Figure \ref{fig:t_off=1000} (with the first panel, $t_{\rm kin}$ = 400 yr, removed), 0.2 pc from the source along the jet axis, one third of the way from the source to the edge of the simulation.  We can see that as the jet propagates through the medium, the bow shock entrains the surroundings, clearing out a cavity around the jet.  The density of the jet was allowed to vary as a free parameter based on the ambient density and the force imparted by the central source. The jet itself appears to have a density of 4$\times 10^4$ cm$^{-3}$, with the swept up surrounding material having a much lower, but non-zero, density. As the simulation progresses, Figure \ref{fig:dens} shows that the bow shock propagates further from the jet axis, causing the outflow to lose collimation (at that point) with time.

\section{Discussion}
\label{sec:discussion}

To determine the inclination angle of the outflow in G5.89, we compared our observations to the models of Cabrit \& Bertout (1990), as well as our own simulations.  In both cases, we found the inclination angle to be $\sim 45^{\circ}$.  We can apply this correction factor to our derived outflow properties (as was done in Table \ref{tab:results}), as well as the total extent of the outflow.  Correcting for the inclination angle changes the total extent of the observed outflow to be 1.6 pc.  Applying the same correction to the simulated outflow, the extent of the outflows in the $t_{\rm off}$ = 1000 yr and $\infty$ cases is 1.55 pc, similar to that observed.  The same is not true of the shorter jet lifetime simulations.  Even with inclination angle effects taken into consideration, the extent of the outflows in the $t_{\rm off}$ = 300 yr and 500 yr cases (respectively) were 1 and 1.3 pc, too small to match observations.

From our simulations, the only distinguishing characteristic between the $t_{\rm off}$ = 1000 yr and $\infty$ cases is the length of the jet, which we cannot observationally constrain. In the $t_{\rm off}$ = 1000 yr case, the jet is only visible towards the edges of the flow, whereas in the $t_{\rm off} = \infty$ case, the jet is visible from the source to the edge.   It is interesting to note that the jets with lifetimes of at least 1000 yr both appear to reach the same extent within 2000 yr, the kinematic age of the outflow.  It is unclear why this happens, however we suggest it may be a result of the chosen ambient density. At lower densities, the jet would be able to easily entrain the surrounding gas, while maintaining a high enough velocity to reach  larger sizes on much shorter timescales.  Conversely, with higher ambient densities, the jet transfers its momentum to the outflow and cannot reach the extent of the observed G5.89 outflow. This issue will be addressed in a future paper.

A comparison of our observations with our simulations suggests a jet lifetime of $\gtrsim$ 1000 yr.  This could indicate that the large scale molecular outflow in G5.89 is a remnant of previous star formation activity.  If we use the accretion to outflow rate ratio discussed earlier (10; Hartmann, 2000), in order for 60 M$_{\odot}$ to buildup requires 2000 yrs.  However, if we increase the ratio to 14 (a rough upper limit suggested by Beuther et al. 2002), the outflow could conceivably be powered by a jet active for as little as 1500 yr.  Since jet activity is linked to accretion, this would give an accretion timescale of only 1500 yr.

Some theories for the formation of UCHII regions suggest that they can only form once accretion has fallen below a critical value (e.g. Yorke 1984, Garay \& Lizano 1999, Yorke et al. 2002).  The UCHII region surrounding G5.89 has been previously shown to be 600 yr old (Acord et al. 1998, Zijlstra \& Pottasch 1988), and physically linked to the source of the outflow (Sollins et al. 2004).    If these results are correct, and accretion must halt before the formation of the UCHII region, an accretion timescale of 1500 yr appears to fit with the age of the UCHII region and kinematic age of the outflow (2000 yr). 

Using Figure \ref{fig:convol}, we can compare the column density in the observed outflow to that in the simulations.  The H$_2$ column density within the central convolved pixel of the simulations is 2.7$\times10^{20}$ cm$^{-2}$ integrated along the emitting region of the simulation and the 4 km s$^{-1}$ velocity bin centered at 4 km s$^{-1}$.  While this portion of the observed red shifted CO is self absorbed, we can use the equivalent velocity bin on the blue shifted side, as the simulations should be symmetric.  Using the integrated intensity of that CO bin (using the same method described in Section \ref{sec:observed_properties}), the CO column density is $\sim 1\times10^{17}$ cm $^{-2}$.  This results in a CO abundance of [CO]/[H${_2}$] = 3.7$\times10^{-4}$, which is approximately within a factor of two of the derived abundance for this region from van der Tak et al.(2000), suggesting that the observations and simulations are well matched.

Other signposts of ongoing star formation include the presence of a protostellar disk, and bullets of molecular gas near the central source detectable in the infrared.  To date, no disk has been detected in G5.89 (Sridharan et al. 2002).  If molecular bullets were released from the central source within the last 500 yr ($t_{\rm kin}$ = 1500 yr), we should be able to detect them within a given radius of the source.  If we assume they would be moving outward at the same rate as the highest velocity outflow (not jet) gas (70 km s$^{-1}$), the farthest they could reach in 500 yr is 0.03 pc.  Any molecular bullets within that radius would have been emitted more recently, and suggest that the jet has been active for longer than 1500 yr.  The field of view of the infrared observations of Feldt et al. (2003) is also approximately 0.03 pc, and no bullets can be seen in their observations, however this region is highly extincted.

Given the results of our observations and simulations, and lack of evidence to the contrary, it is plausible to suggest that the outflow in G5.89 is what remains of a now extinct jet.    The results suggested here have interesting implications for the study of massive star formation. It is possible that accretion timescales are much shorter than previously believed (as little as 1000 yr).  It may also be possible that the apparent  association between massive outflows and UCHII regions may be an artifact of the short massive star formation timescale; that an observed remnant outflow may not be indicative of ongoing accretion.  While others have suggested that infall can continue through a Hypercompact HII region (e.g. Keto 2003), we suggest that this is not the case for G5.89.

\section{Conclusions}
\label{sec:conclusions}

We have created the largest map of the G5.89 outflow to date, and have, for the first time, captured the full extent of the large scale outflow. Through comparison of simulated and observed channel maps, we suggest the outflow is inclined by $\sim 45^{\circ}$ to the line of sight.  This implies that the combined extent of the two outflow lobes is 1.6 pc.  From our observations, we were also able to constrain the mass, momentum, energy, etc. of the outflowing gas. 

We used these observationally derived parameters as input constraints on a set of MHD simulations of a jet entrained molecular outflow.  Our results suggest only one source powered the outflow, and that the jet must have been active for a minimum of 1000 yr to account for the spatial extent of the observed outflow.  This jet activity timescale is consistent with a 600 yr old UCHII region, and a 2000 year old outflow.

\acknowledgments
The authors would like to thank T.K. Sridharan for helpful discussions, and the anonymous referee for strengthening the paper. We would also like to acknowledge the National Science and Engineering Research Council of Canada for their financial support.

\clearpage

\begin{table}
\begin{center}
\begin{tabular}{rcclcclccccc}
\hline \hline
\multicolumn{3}{c}{Tracer} & &\multicolumn{2}{c}{Map Parameters} & &  \multicolumn{5}{c}{System Parameters}\\
\cline{1-3} \cline{5-6} \cline{8-12}\\
\multicolumn{2}{c}{Transition} & $\nu$ & & Mapsize & Spacing &&  $T_{sys}$ & $T_{rms}$ & $\tau$ & $\Delta v$ & $t_{int}$ \\
\hline
 & &  (GHz) & & (Pixels) &  ($''$) & &(K) & (K) & & (km/s) & (s)\\
 $^{12}$CO & \trans{3}{2} & 345.796 & & 28$\times$17 & 7  & & 507.9 &0.241 & 0.20 & 1.08 & 2.57\\
 $^{13}$CO & \trans{3}{2} & 330.588 & & 15$\times$15 & 7  & & 759.2 &0.190 & 0.37  & 1.13 & 1.27\\
 SiO    & \trans{8}{7} & 347.331 & & 12$\times$12 & 7  & & 660.4 &0.059 & 0.26 & 1.08 & 0.97\\
 SiO    & \trans{5}{4} & 217.105 & & 8$\times$8 & 10 & & 467.3 &0.190 & 0.15 & 3.45 & 1.82\\
 \hline \hline
 \end{tabular}
 \caption[Observations Parameters] {Observation Parameters.  System parameters (system temperature, rms noise limit, atmospheric opacity, velocity resolution and integration times) are given for the central position in each map.}
 \label{tab:obs_param}
 \end{center}
 \end{table}

\begin{table}
\begin{center}
\begin{tabular}{ll|rr}
\hline
\hline
& & \multicolumn{2}{c}{Outflow Value}\\
Characteristic & Equation & Red & blue\\
\hline
Mass ($M_\odot$)& $M = N_{H_2}\cdot A\cdot m_{H_2}/M_{\odot}$ &                 0.6 & 2.7\\
Momentum ($M_{\odot}$ km s$^{-1}$) & $P = \sum_i (m_i/M_{\odot})|v_i-v_{LSR}|$          &17 & 79\\
Energy ($\times10^{45}$ erg) & $E = 1/2\sum_i m_i(v_i-v_{LSR})^2$ &             5.6& 29\\
Luminosity (L$_{\odot}$) & $L_{mech} = E/t$ &                                   23 &118 \\
Mass loss ($\times 10^{-3} M_{\odot} yr^{-1}$) & $\dot{M} = M/t$ &              0.3&1.4 \\
Force ($\times10 ^{-3}M_{\odot}$ km s$^{-1}$ yr$^{-1}$) & $\dot{P} = P/t$ &     8.5&59 \\
\hline
\hline
\end{tabular}
\caption{Flow characteristics corrected for an inclination angle of 45$^{\circ}$.  The $t$ in the mass loss rate and outflow force equations refers to the
 kinematic age of the outflow derived from observations ($t_{\rm kin}$ = 2000 yr).}
\label{tab:results}
\end{center}
\end{table}

 \begin{table}[hbt!]
 \begin{center}
 \begin{tabular}{rl|lc}
 \hline \hline
\multicolumn{4}{c}{Simulation initial conditions}\\
\hline
 \multicolumn{2}{c|}{Source Parameters} & \multicolumn{2}{c}{Ambient Constraints} \\
  \hline
 $M_*$ & 60 M$_{\odot}$                         & $T_{\rm{ism}}$ & 100 K\\
 $R_*$ & 12 $R_{\odot}$                         & $n_{\rm{ism}}$ & $10^4$ cm$^{-3}$\\
 $v_w$ & 552 km s$^{-1}$                        & $c_s$ & 1 km s$^{-1}$\\ 
 $\beta$ & 10$^{-4}$              \\
 \hline
 \hline
 \end{tabular}
 \caption{Constraints on values for source and ambient parameters.} \label{tab:init_sim}
 \end{center}
 \end{table}

\clearpage

\begin{figure}[hbt!]
\includegraphics[scale=0.55,angle=-90]{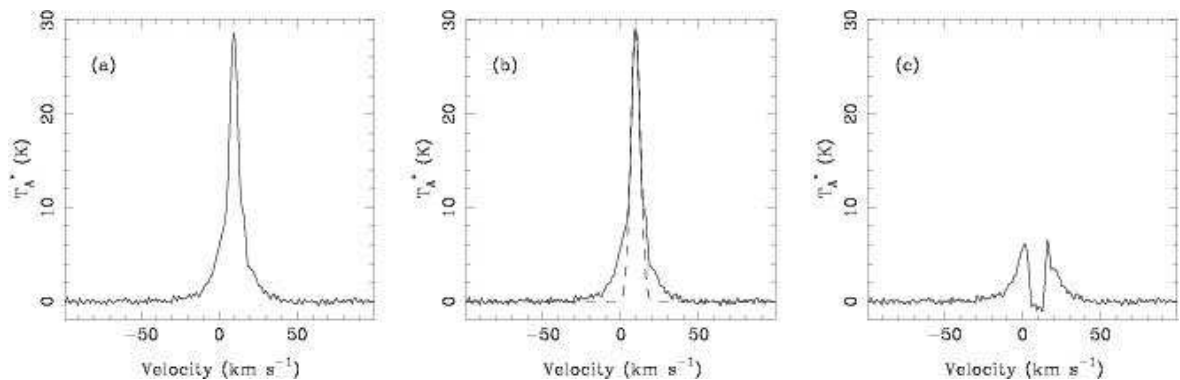}
\caption{Gaussian fit removal from the $^{13}$CO spectrum at our map center. {\bf(a)}: Original $^{13}$CO spectrum from our map. {\bf(b)}: Gaussian fit (dashed line) and $^{13}$CO spectrum. {\bf(c)}: residual $^{13}$CO emission once the Gaussian profile was removed using the RESIDUAL command in CLASS.}
\label{fig:residual}
\end{figure}

\begin{figure}[hbt!]
\vspace{8cm}
\includegraphics{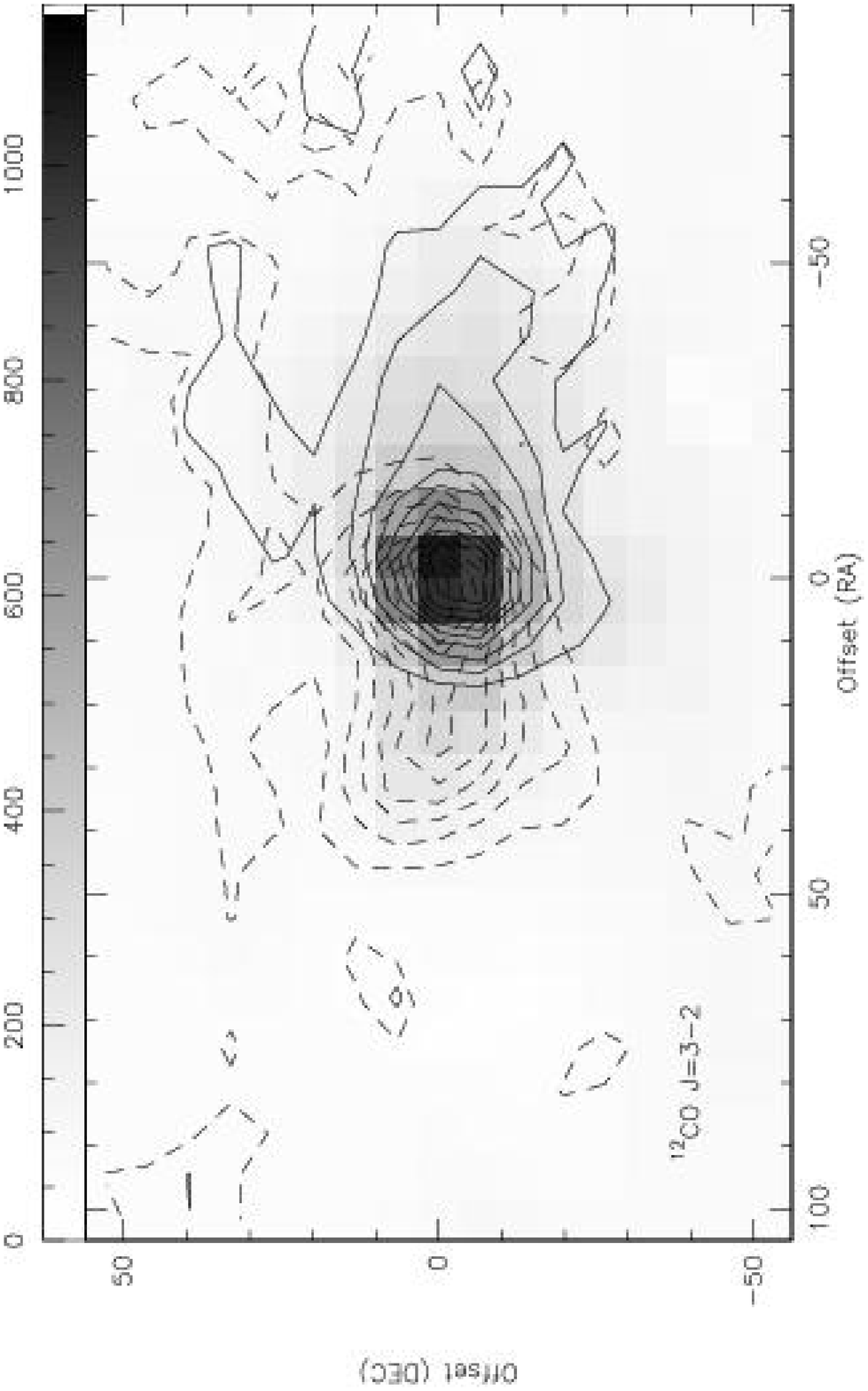}
\includegraphics{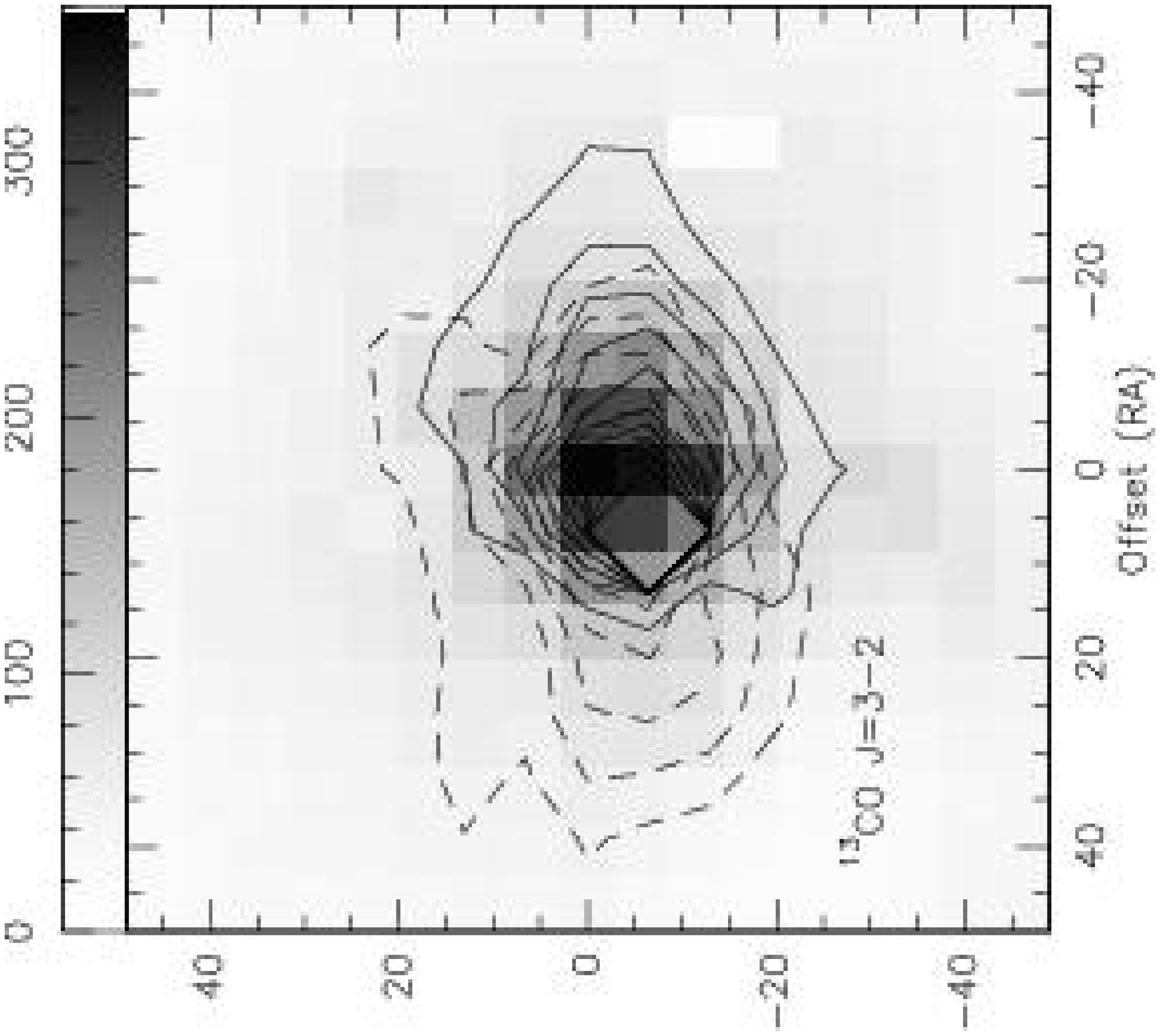}
\caption{$^{12}$CO (left) and $^{13}$CO (right) integrated intensity maps overlaid by red (dashed) and blue (solid) line wing emission.  Contours start at 10\% of the integrated line wing intensity and increase in 10\% increments. The grey scale shows total integrated intensity in units of K Km s$^{-1}$.}
\label{fig:12COmap}
\end{figure}

 \begin{figure}[hbt!]
\includegraphics[scale=0.5,angle=-90]{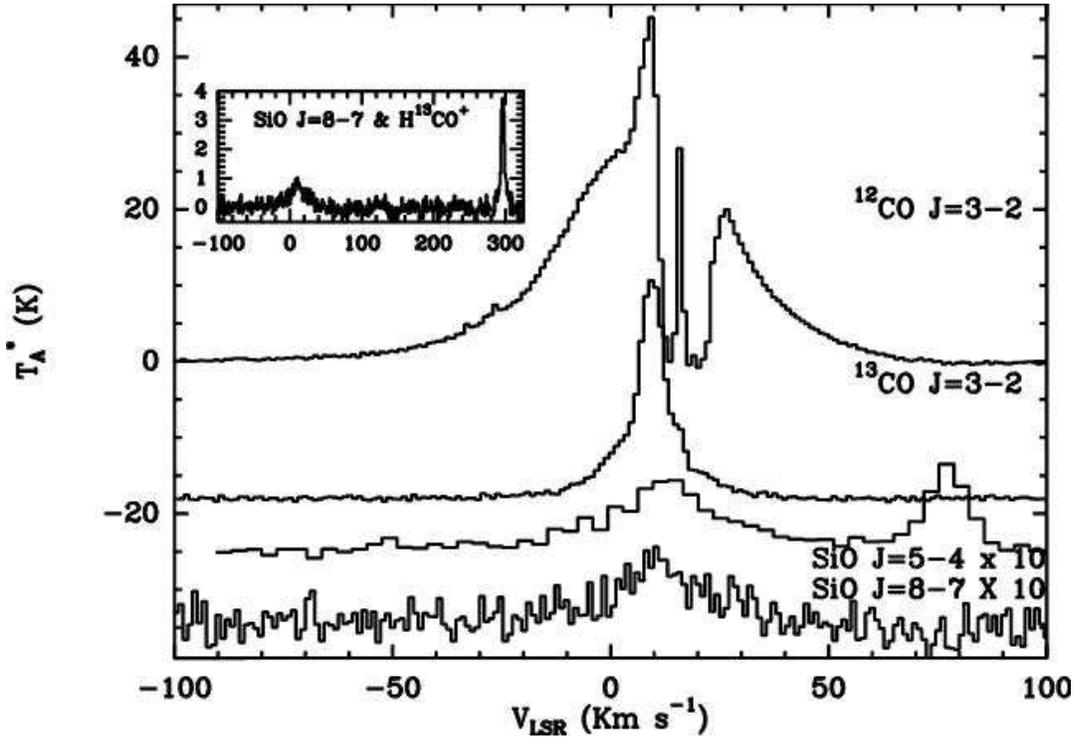}
 \caption[spectra at the central position]{Spectra from the central position in each observed $^{12}$CO, $^{13}$CO, and SiO transition. Note that the $^{13}$CO \trans{3}{2} and SiO \trans{5}{4} and \trans{8}{7} spectra are offset from the $^{12}$CO by 18 K, 25 K, and 35 K (respectively).  The redshifted material in the $^{12}$CO observations is absorbed at two different sets of velocities by cold line of sight clouds.  Based on the observations of HF88, we suggest this is not the case for the $^{13}$CO and SiO.  The SiO spectra have been multiplied by a factor of 10 for comparison with the CO spectra.}
 \label{fig:spectra}
 \end{figure}

 \begin{figure}[hbt!]
\includegraphics[scale=0.5,angle=-90]{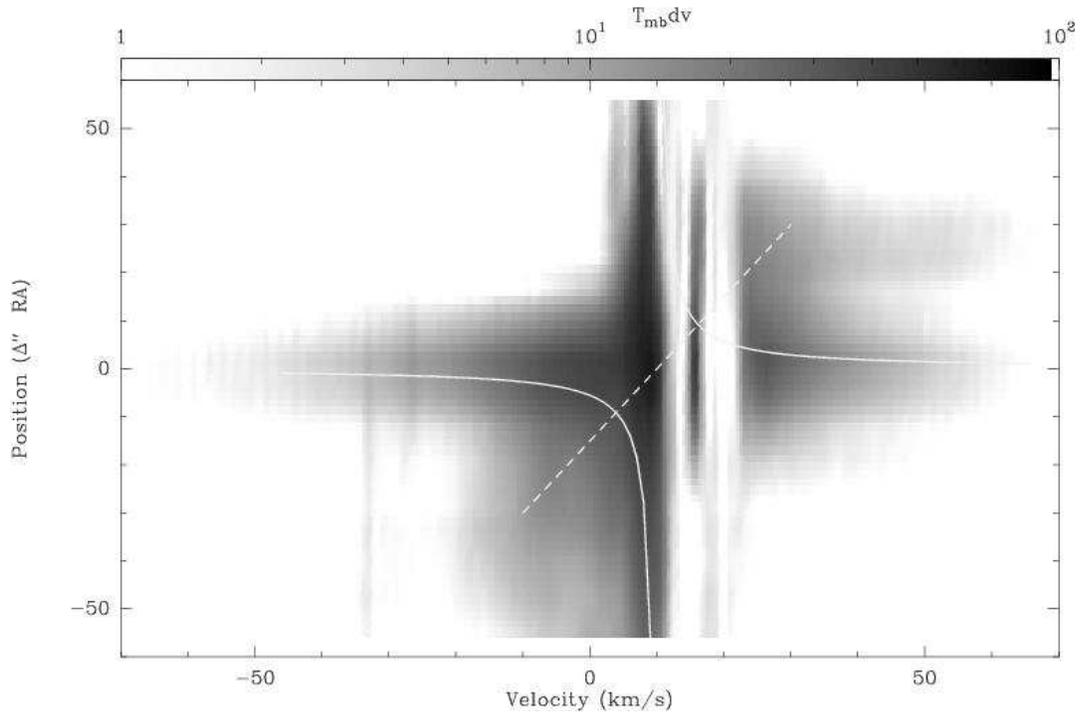}
 \caption[P-V diagram for G5.89]{P-V diagram for G5.89.  The two solid white lines represent contours of $v\propto r^{-1}$, while the dashed diagonal line represents $v\propto r$.  There appears to be evidence for both a larger scale decelerating outflow and smaller scale accelerating outflow. The slice shown above corresponds to $\Delta$ DEC = 0 in Figure \ref{fig:12COmap}.} \label{fig:w28a2PV}
 \end{figure}

 \begin{figure}[hbt!]
 \vspace{8cm}
\includegraphics{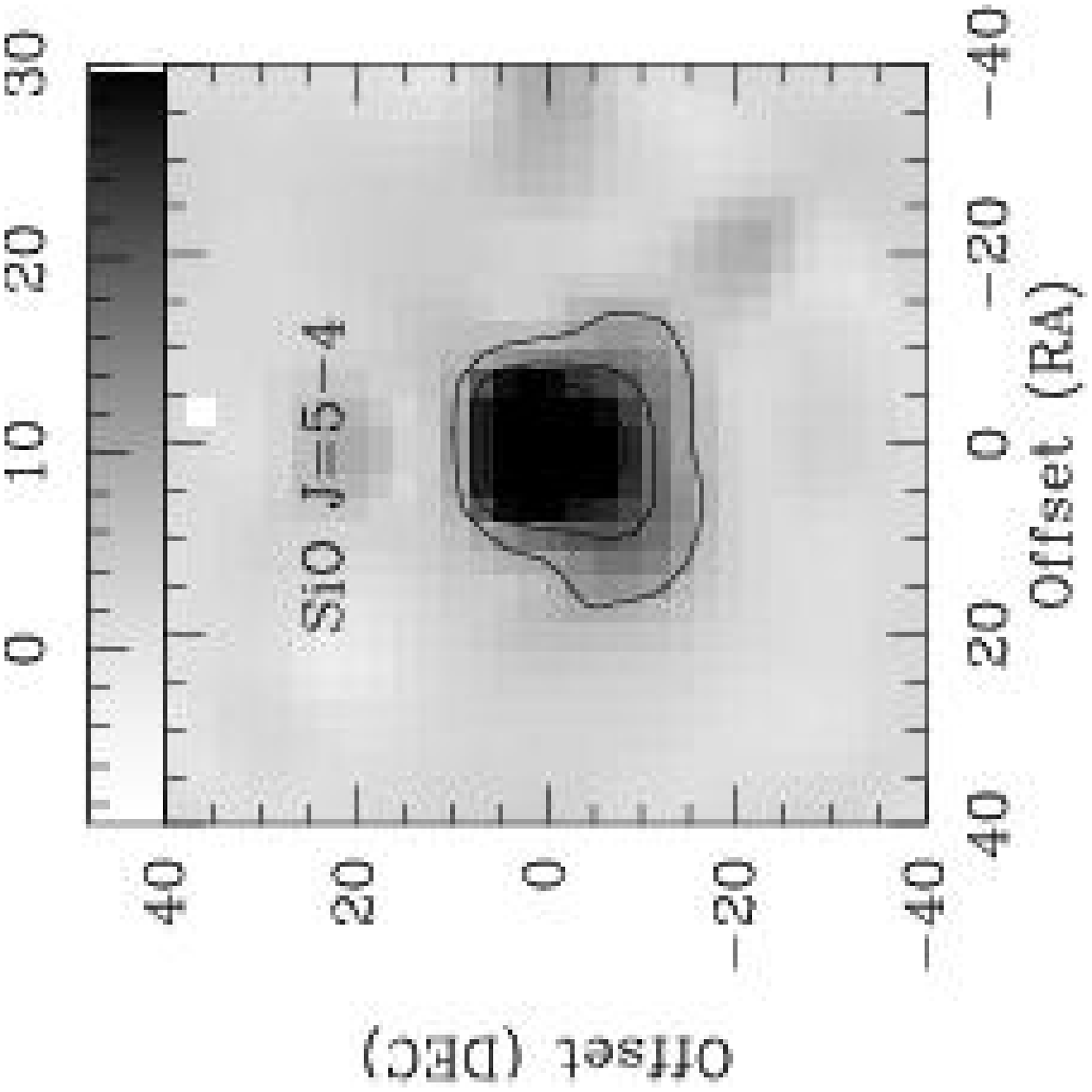}
\includegraphics{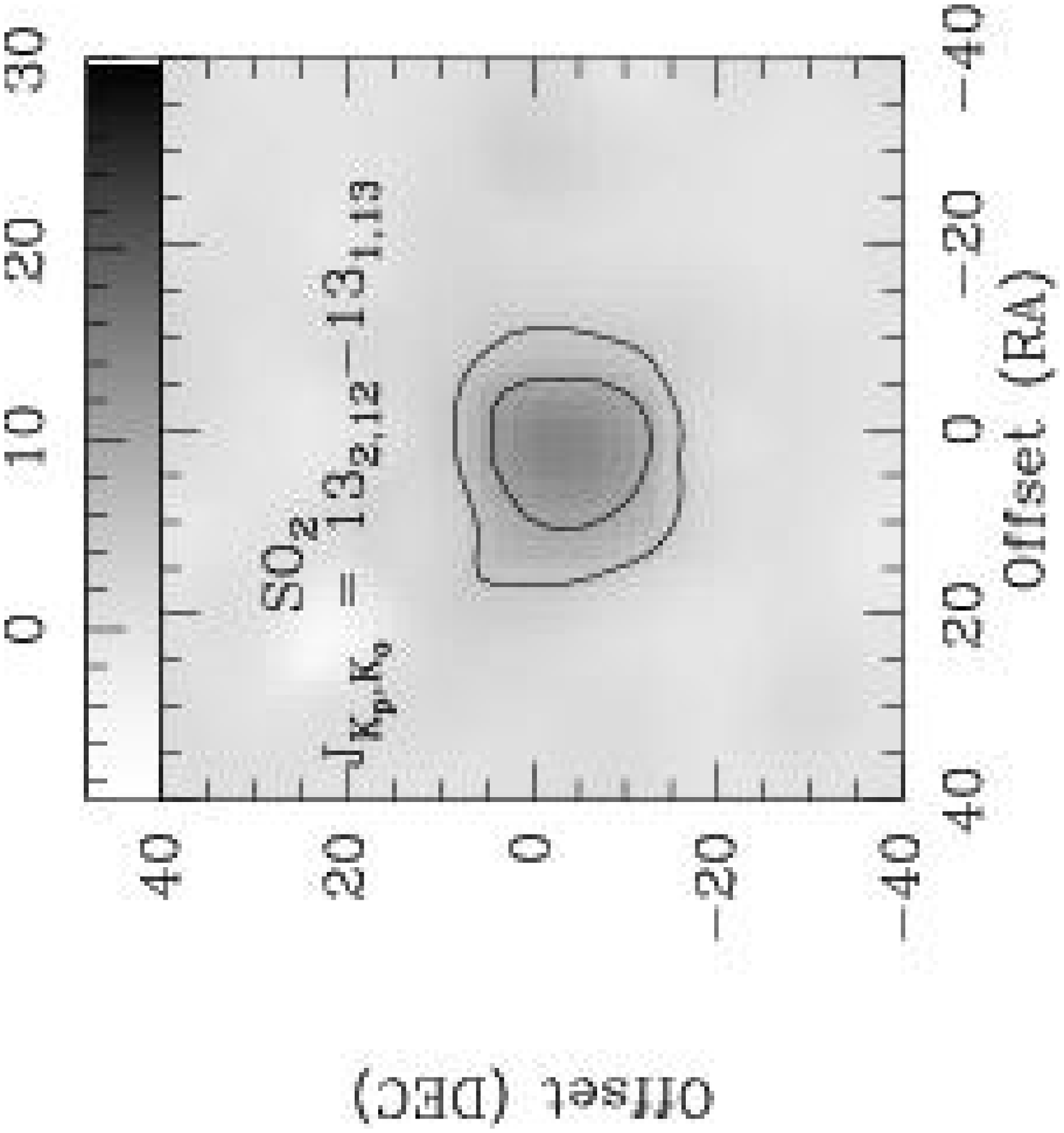}
\includegraphics{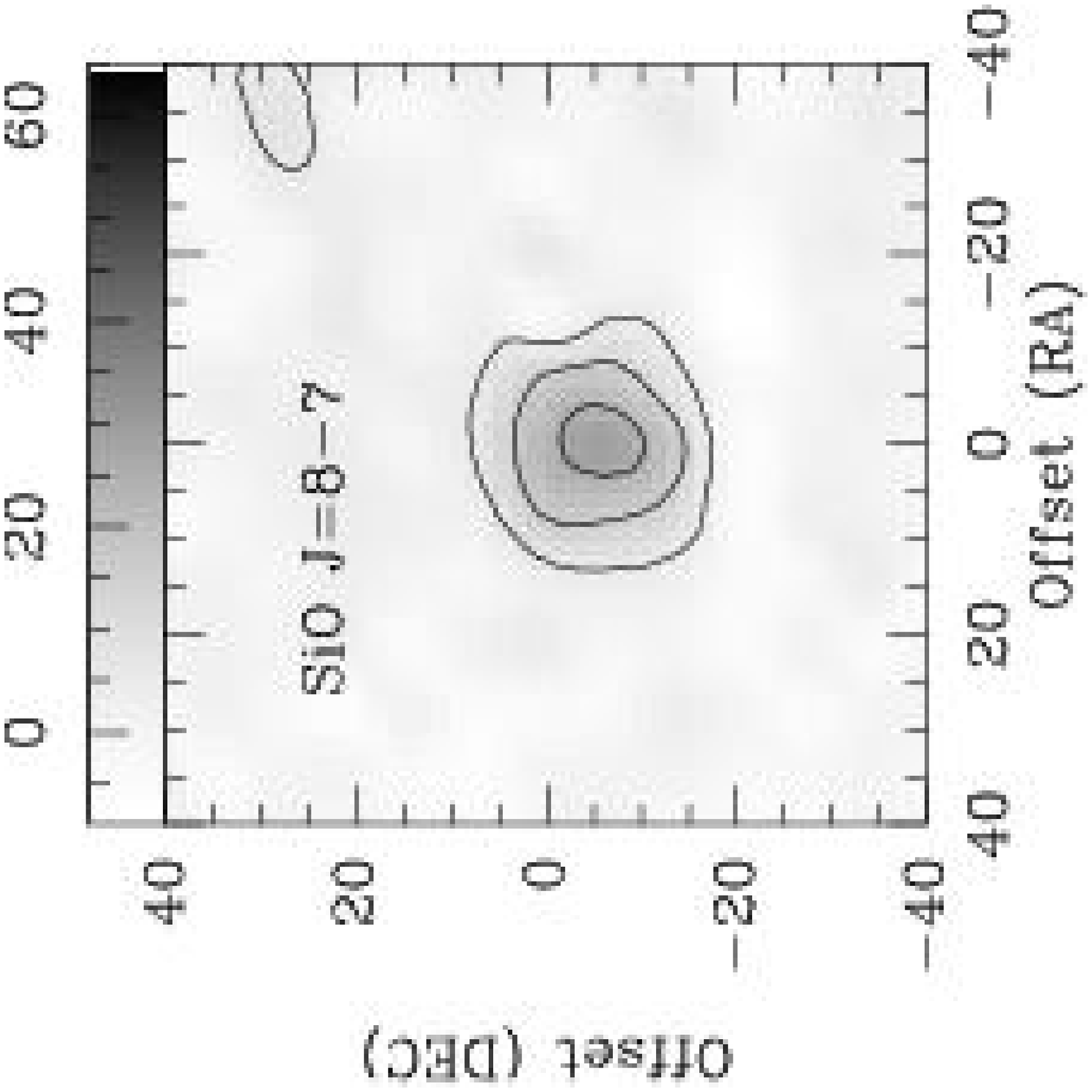}
\includegraphics{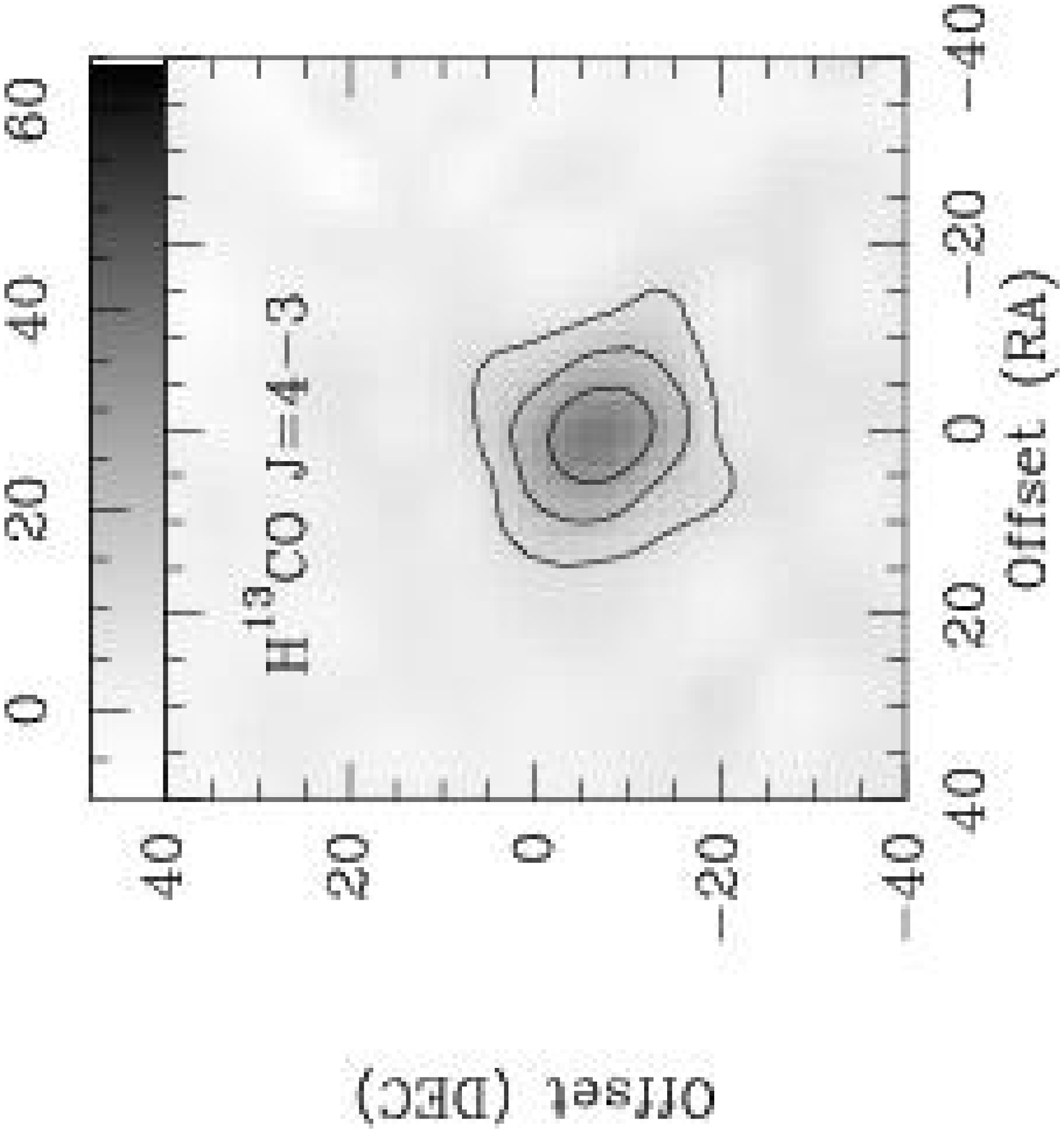}
 \caption[Total Integrated Intensity Maps]{Total integrated intensity maps ($\int T_R^*dv$) for each non CO transition observed at the JCMT.  Specific molecular transitions are given in the upper left hand corner of each map.  The maps are on the same spatial scale, but the integrated intensities represented by the grey scale are as shown at the top of each map (in units of K km s$^{-1}$). For maps of integrated SiO \trans{5}{4}, SO$_2$ \trans{13$_{2,12}$}{13$_{1,13}$}, SiO \trans{8}{7}, and H$^{13}$CO$^+$ \trans{4}{3}, the contours begin at 10, 5, 5, and 10 K km s$^{-1}$ respectively, and increase in steps of 10, 5, 10 and 10 K km s$^{-1}$ respectively.}
 \label{fig:total_intensity}
 \end{figure}

\begin{figure}[hbt!]
\includegraphics[scale=0.5,angle=-90]{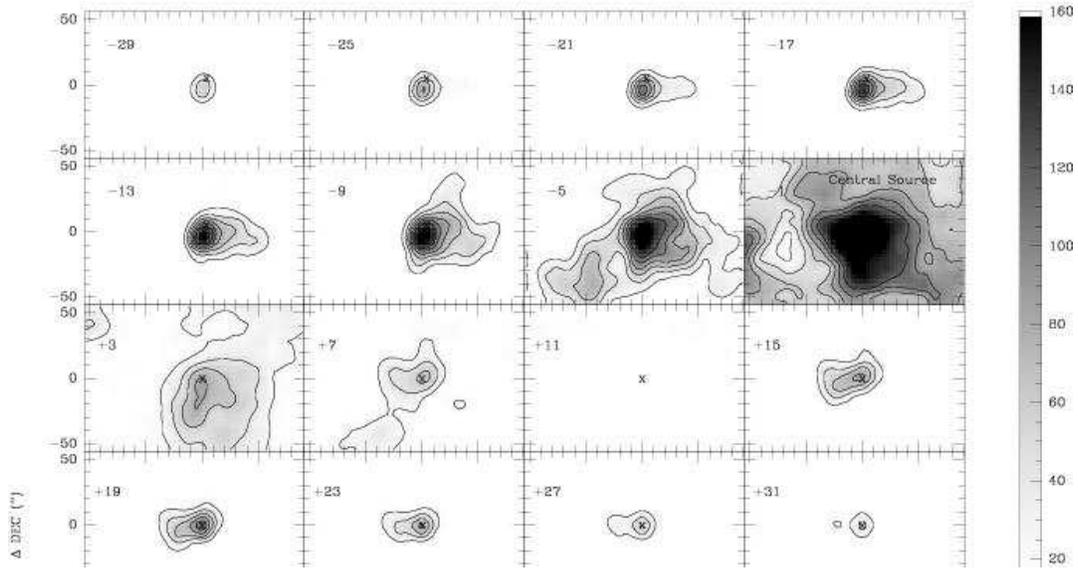}
\caption{Channel maps of the G5.89 outflow in $^{12}$CO \trans{3}{2}.  The number in the upper right hand corner of each frame is the middle velocity of the bin (4 km s$^{-1}$ per bin) with respect to the $V_{LSR}$ of the source (9 km s$^{-1}$).}
\label{fig:channel}
\end{figure}

\begin{figure}[hbt!]
\vspace{7cm}
\begin{center}
\includegraphics{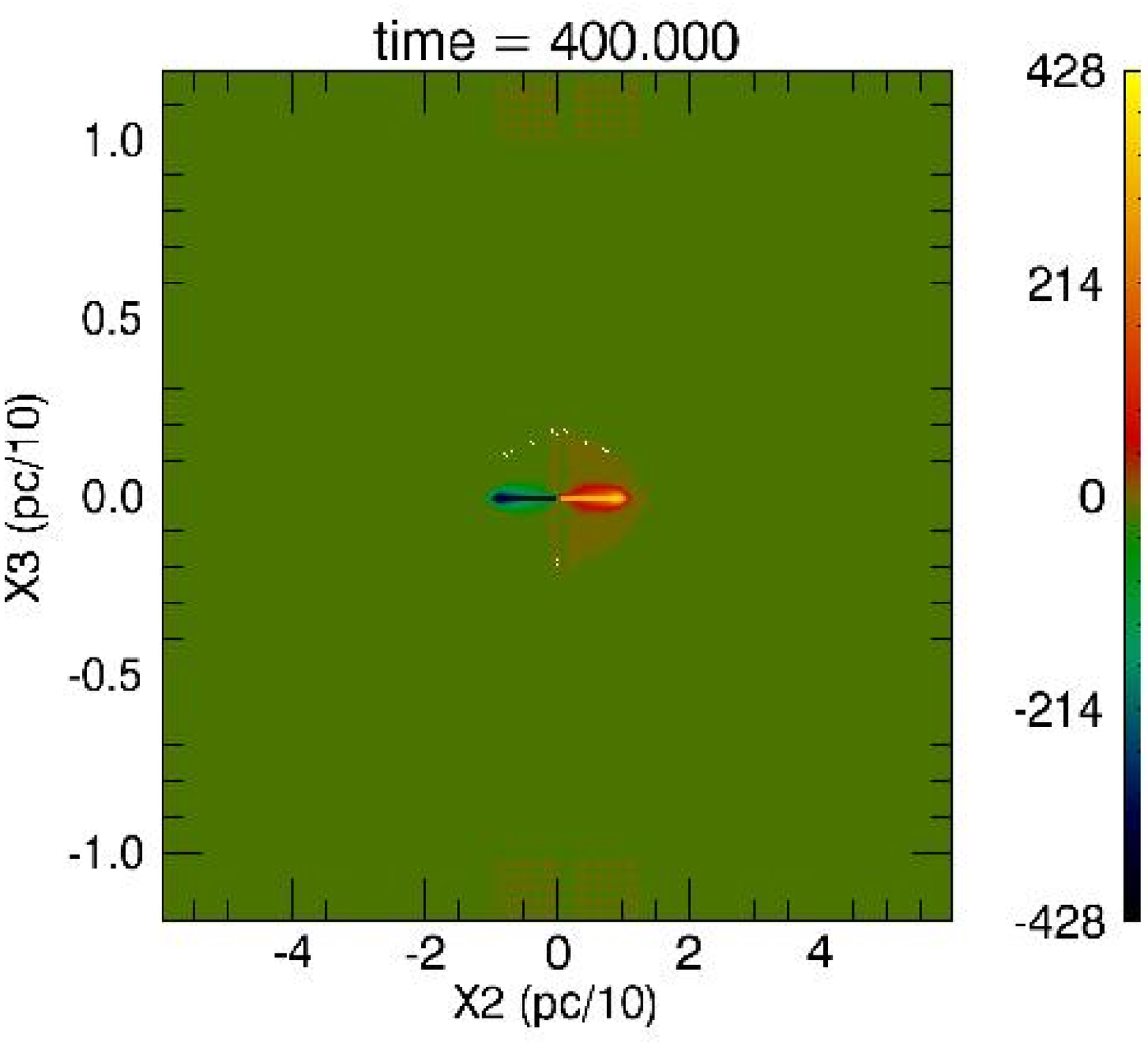}
\includegraphics{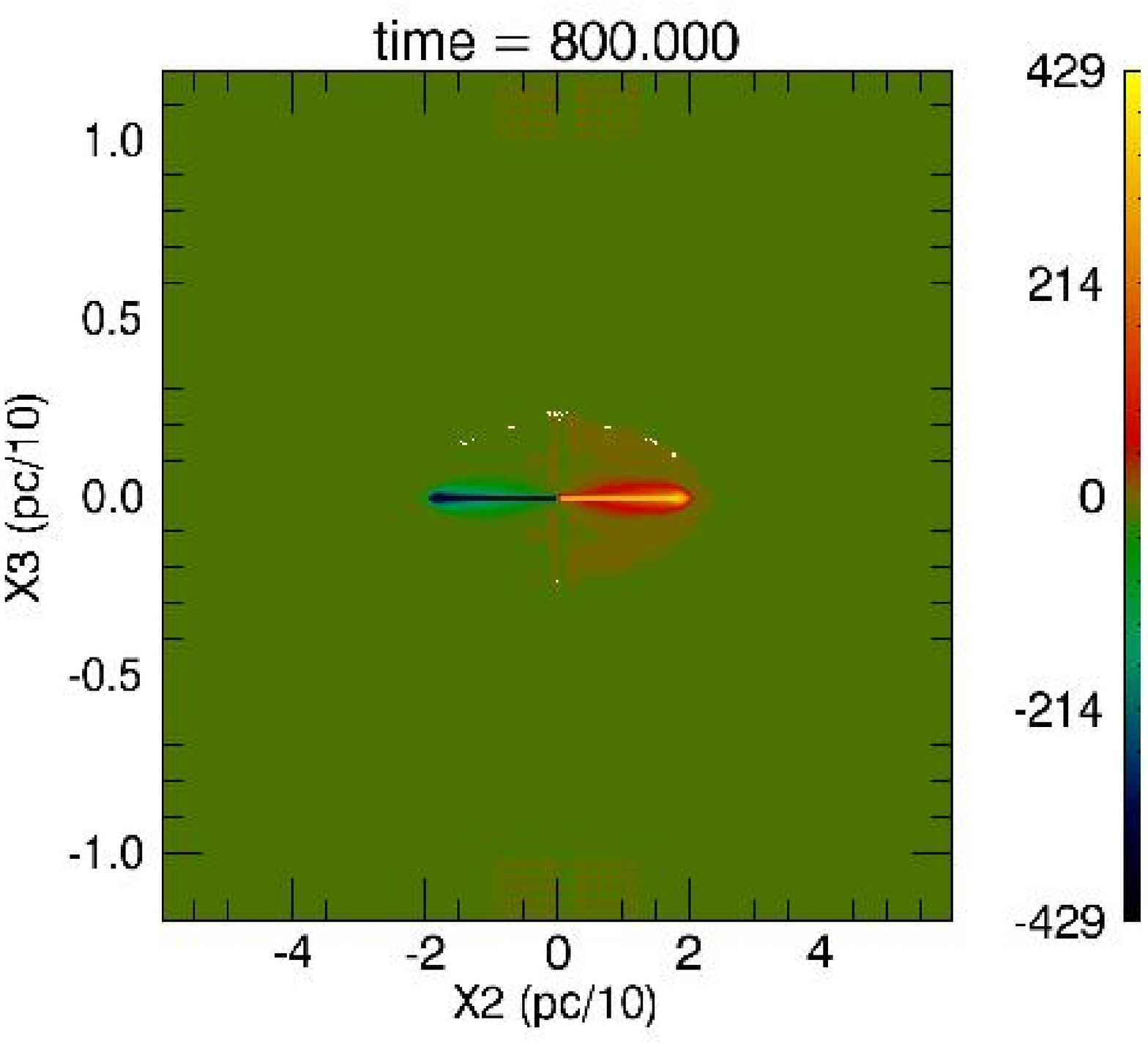}
\includegraphics{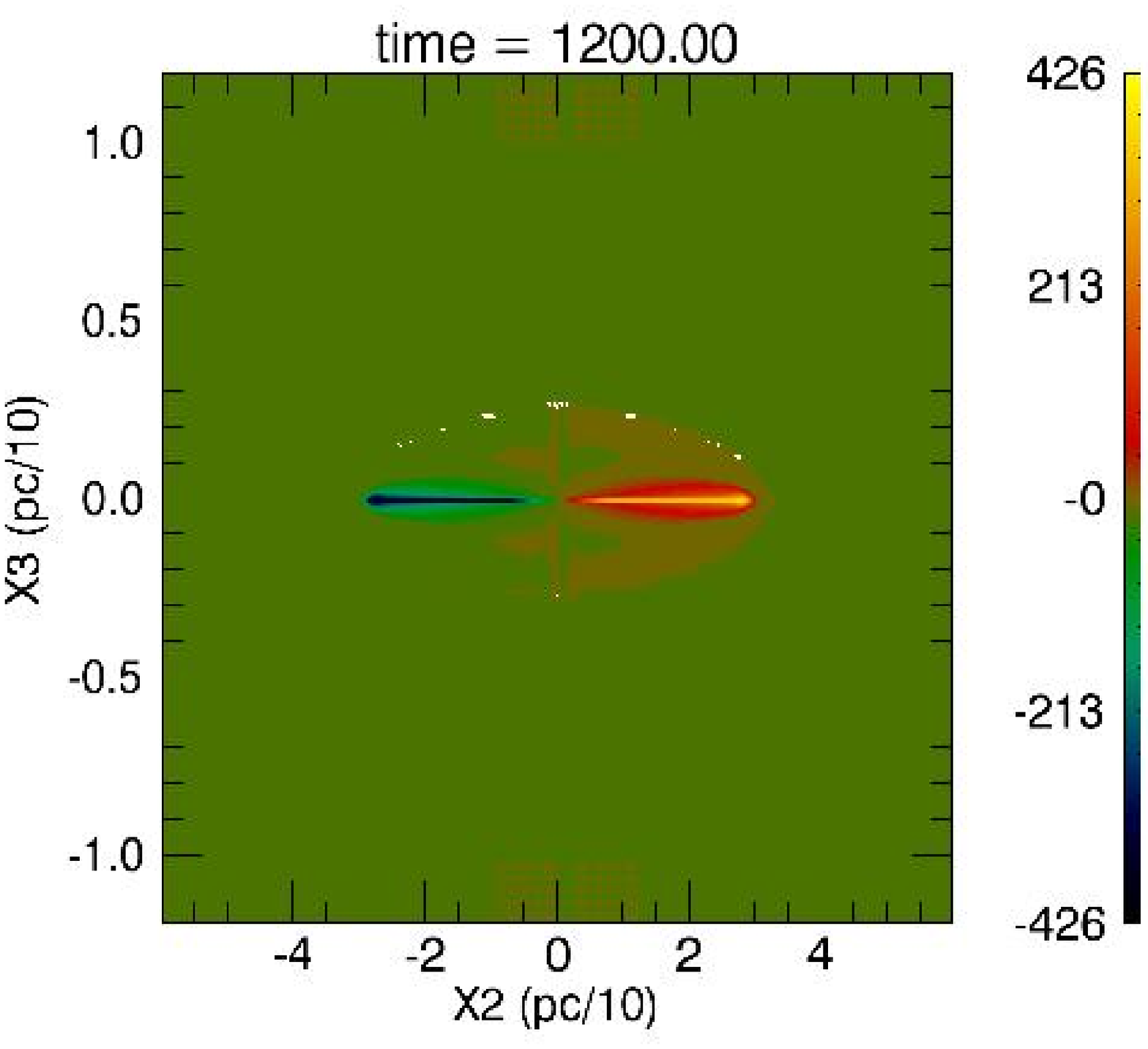}
\includegraphics{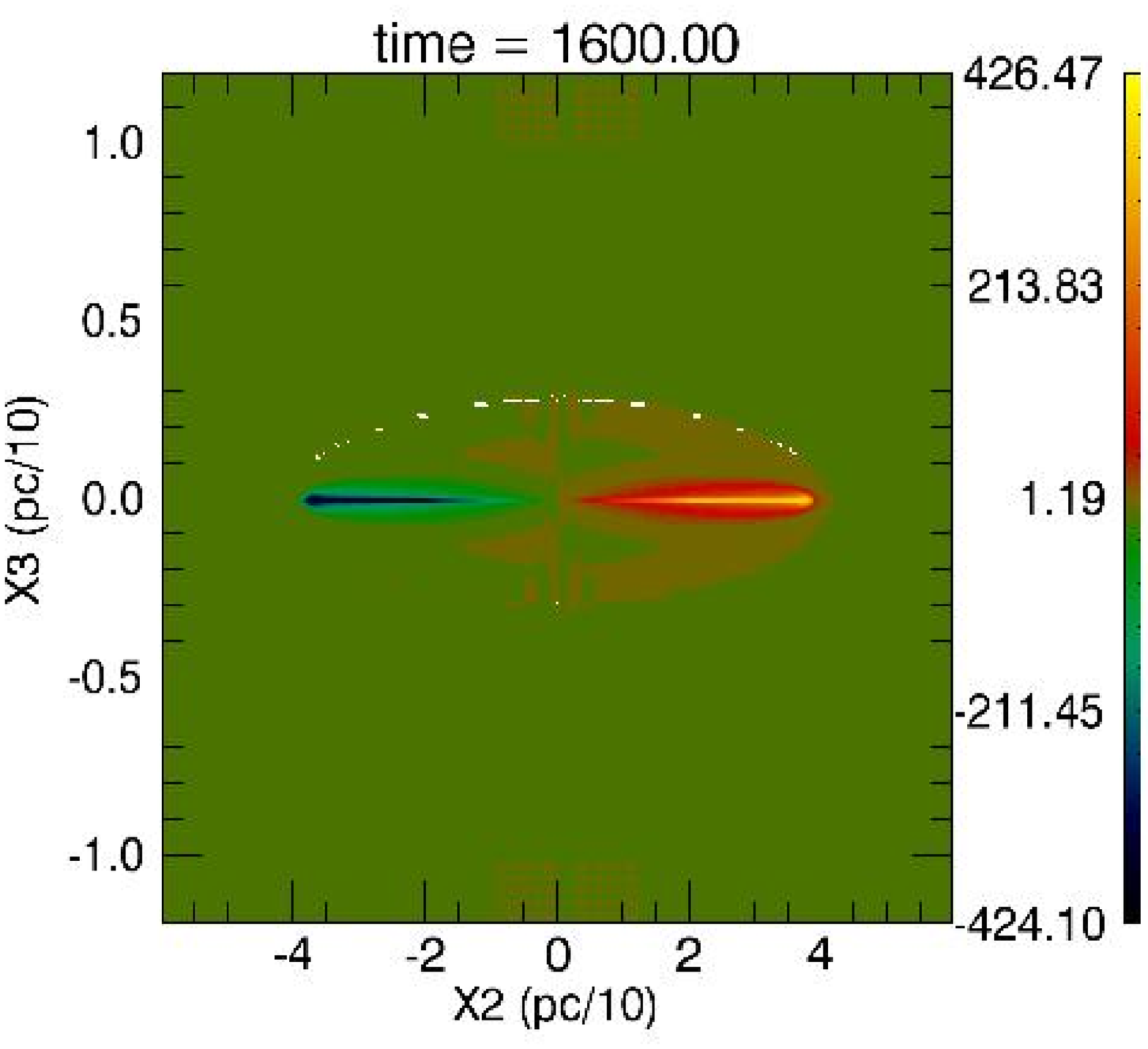}
\includegraphics{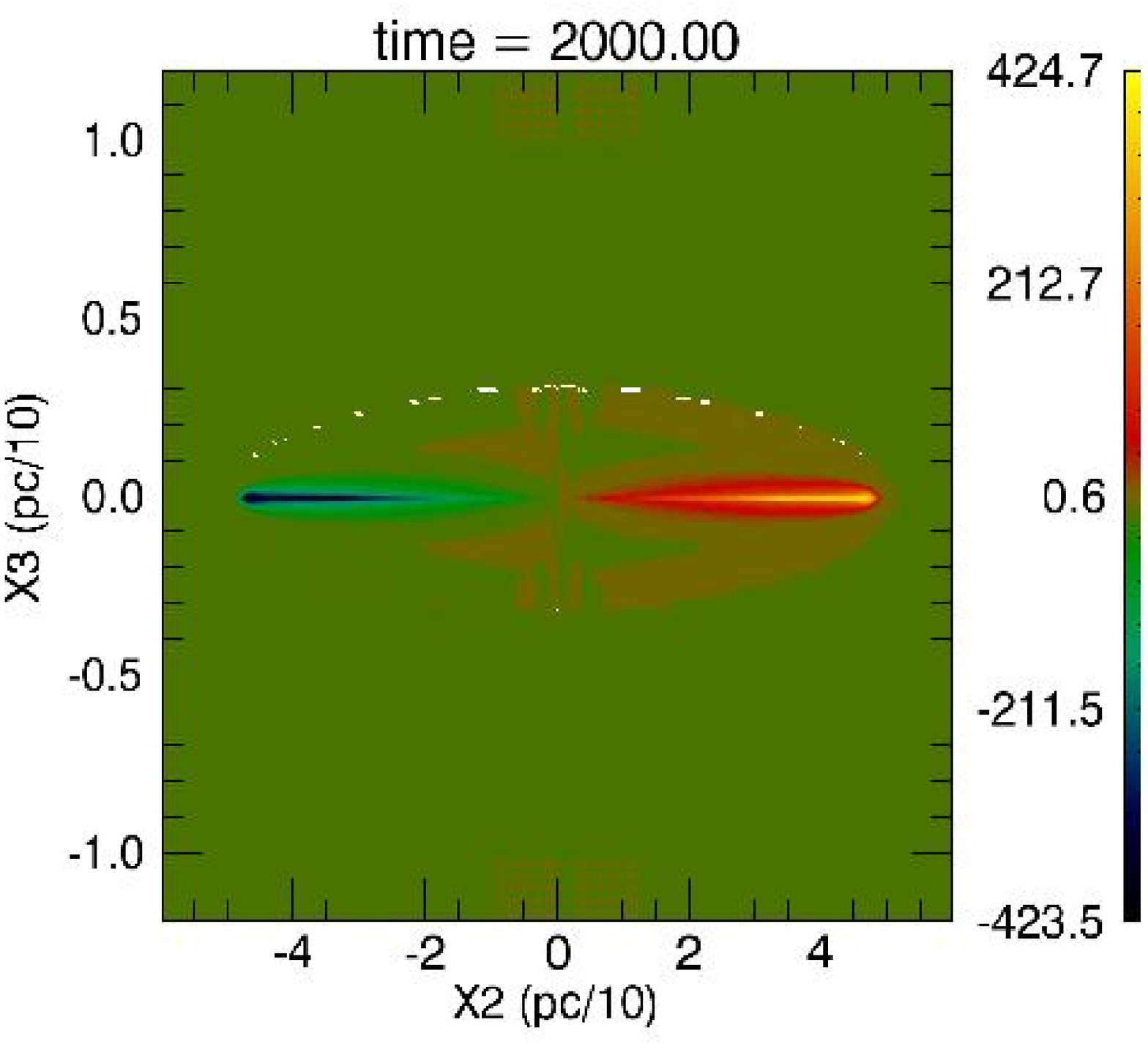}
\includegraphics{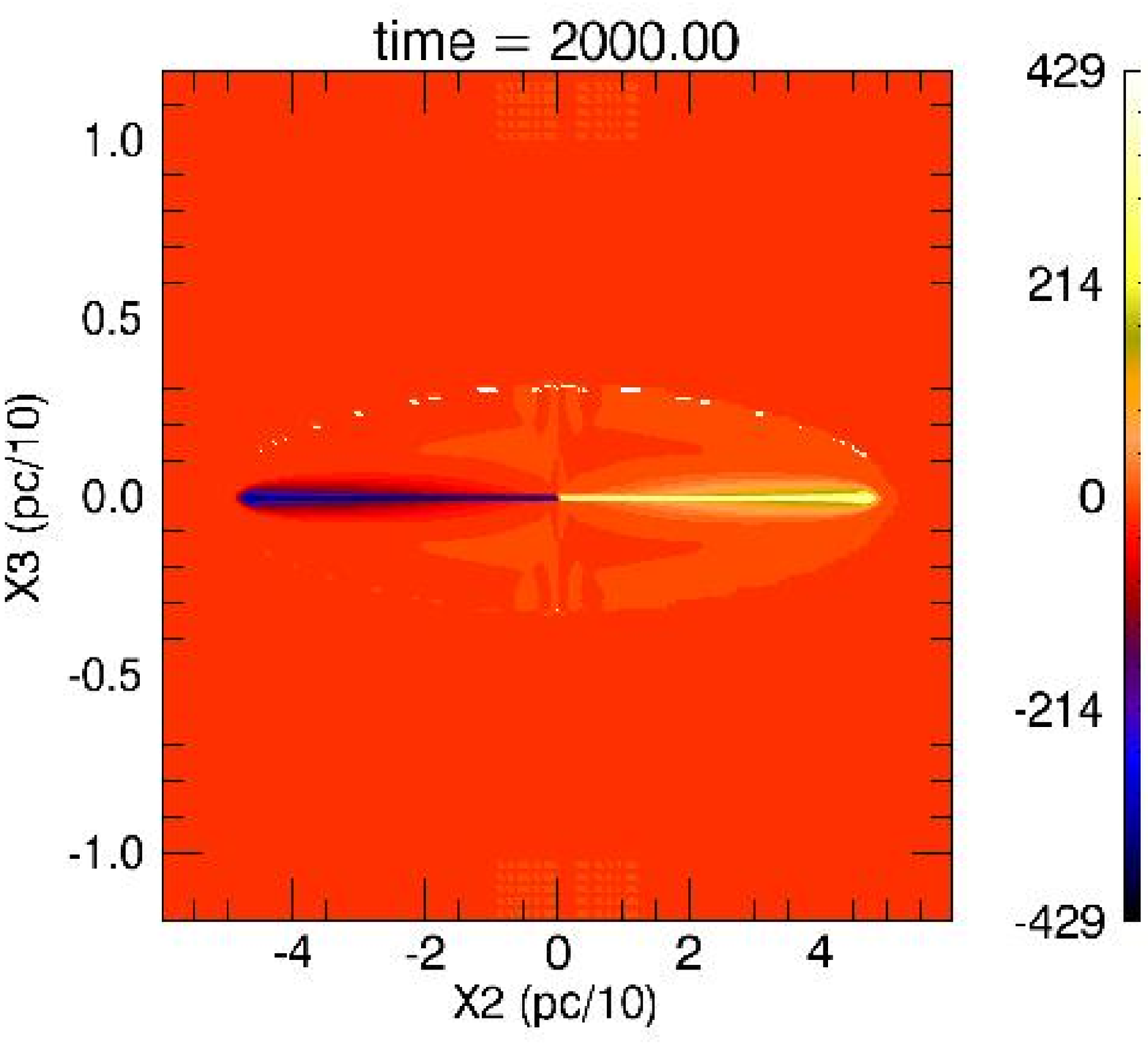}
\end{center}
\caption{The first five panels (with green backgrounds) show velocity contours, and the evolution of the simulated outflow in which the jet was shut off after 1000 years.  Panels show 400 year intervals, starting at 400 years and ending at 2000 years.  The bottom right panel (with red background) shows the $t_{\rm off}$ = 2000 yr outflow at a kinematic age of 2000 yrs for comparison to the $t_{\rm off}$ = 1000 yr case (bottom center).  The extent of the two outflows are the same, however, the jet can be traced back to the central source in the $t_{\rm off}$ = 2000 yr case.  The two dimensional slices of the three dimensional simulations are in plane (or slice) 105 out of 210 slices, through the central source.  The halftone scale represents the velocity of the gas, with the minimum ($\sim$ -430 km s$^{-1}$) and maximum ($\sim$ 430 km s$^{-1}$) velocity given for each panel.  The scales in the $x_2$ and $x_3$ planes are in units of 0.1 pc.}
\label{fig:t_off=1000}
\end{figure}

\begin{figure}[hbt!]
\vspace{18cm}
\begin{center}
\includegraphics{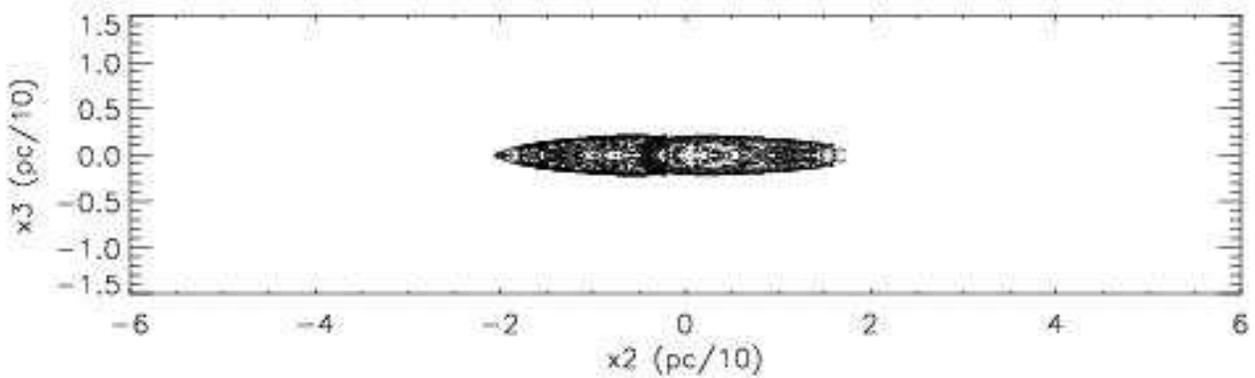}
\includegraphics{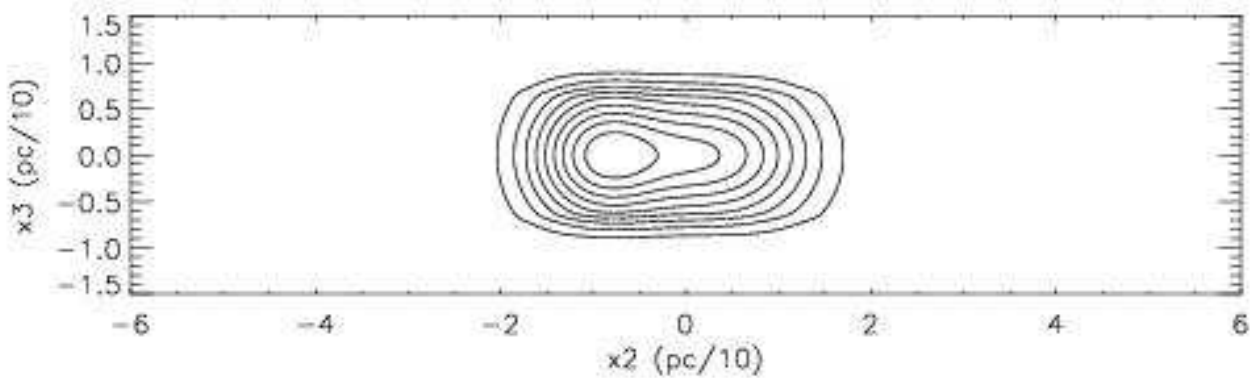}
\end{center}
\caption{Column density contours for simulated jet and outflow, shown at an inclination angle of 45$^{\circ}$. The maps shown are channel maps, taken over the simulated velocity range of 4 km s$^{-1}$ centered at 4 km s$^{-1}$ (e.g. 4$\pm$2 km s$^{-1}$ with respect to the source velocity). The top panel shows the jet, with equal scaling on the x and y axes, while the bottom panel shows the jet convolved with a 0.145 pc (15$''$ at the distance of G5.89) Gaussian for comparison to our JCMT observations (e.g. the panel labelled as ``+3'' in Figure \ref{fig:channel}). There are ten column density contours in each panel, starting from $1.2\times 10^{21}$ cm$^{-2}$ and $2.7\times 10^{20}$ cm$^{-2}$ (for the top and bottom panels respectively) and decreasing in intervals of 10\% of the maximum value. The H$_2$ column density in the central pixel of the lower plot is only a factor of two off from the observed column density.}
\label{fig:convol}
\end{figure}
\clearpage

\begin{figure}[hbt!]
\vspace{14cm}
\begin{center}
\includegraphics{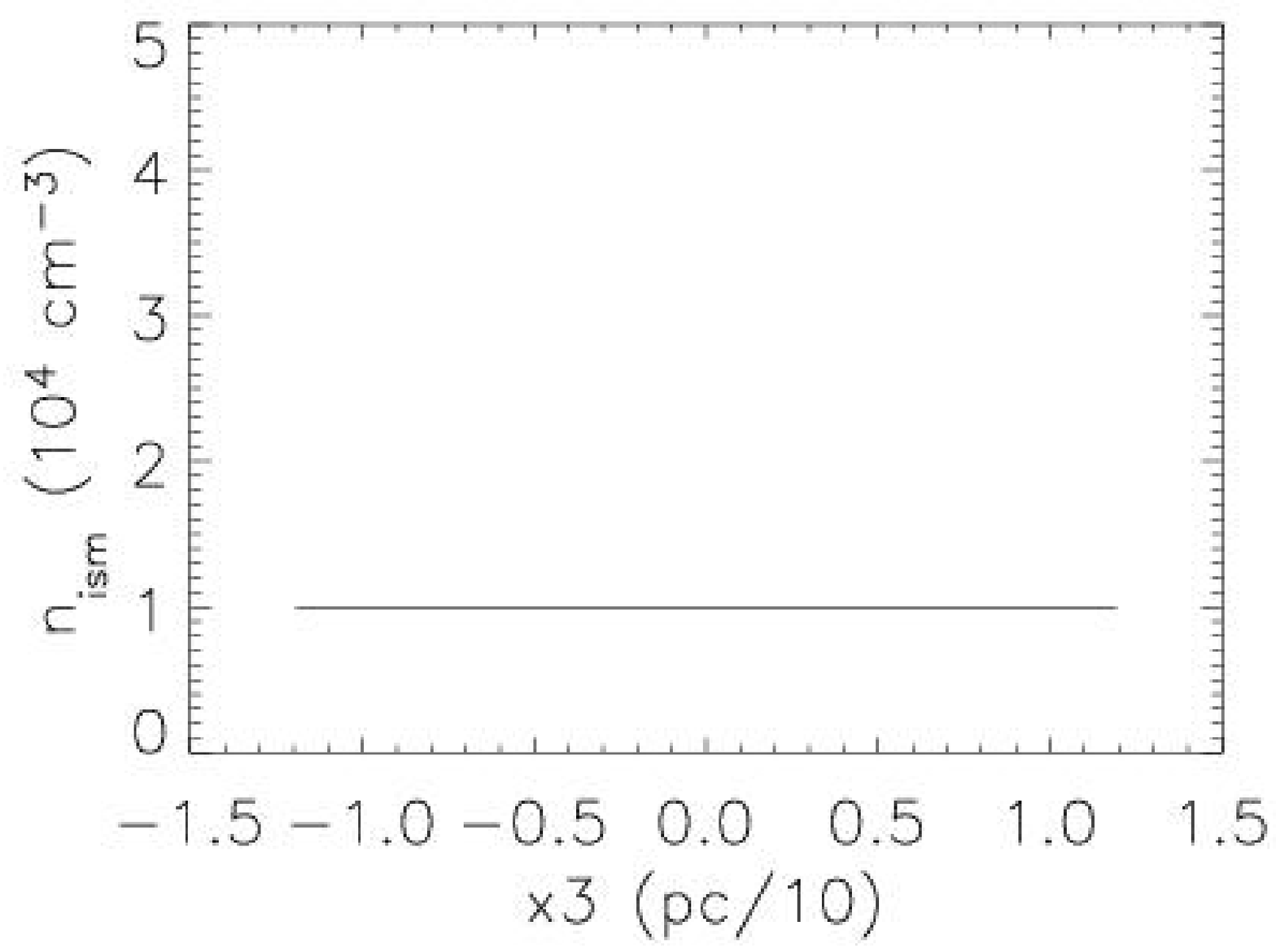}
\includegraphics{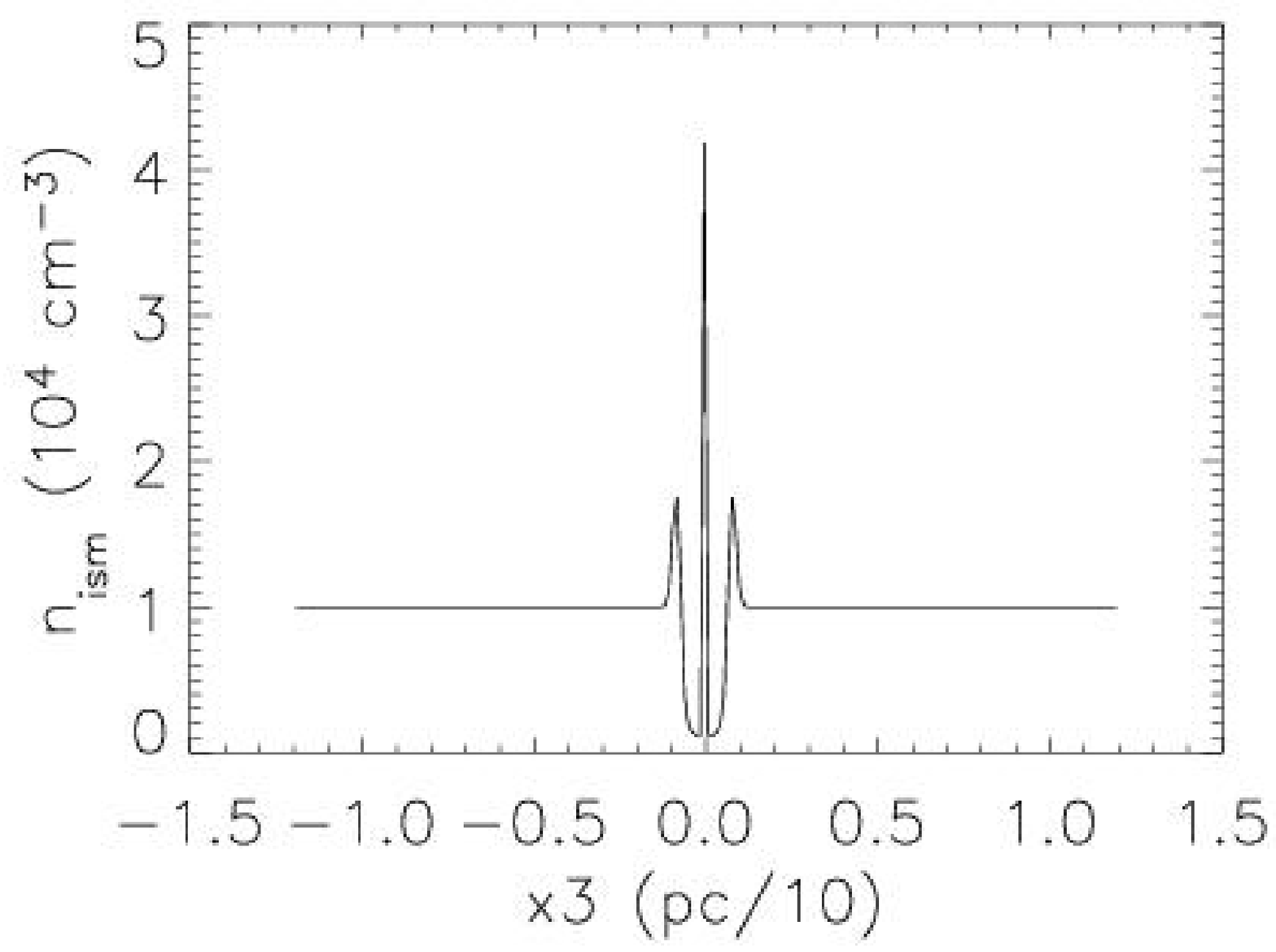}
\includegraphics{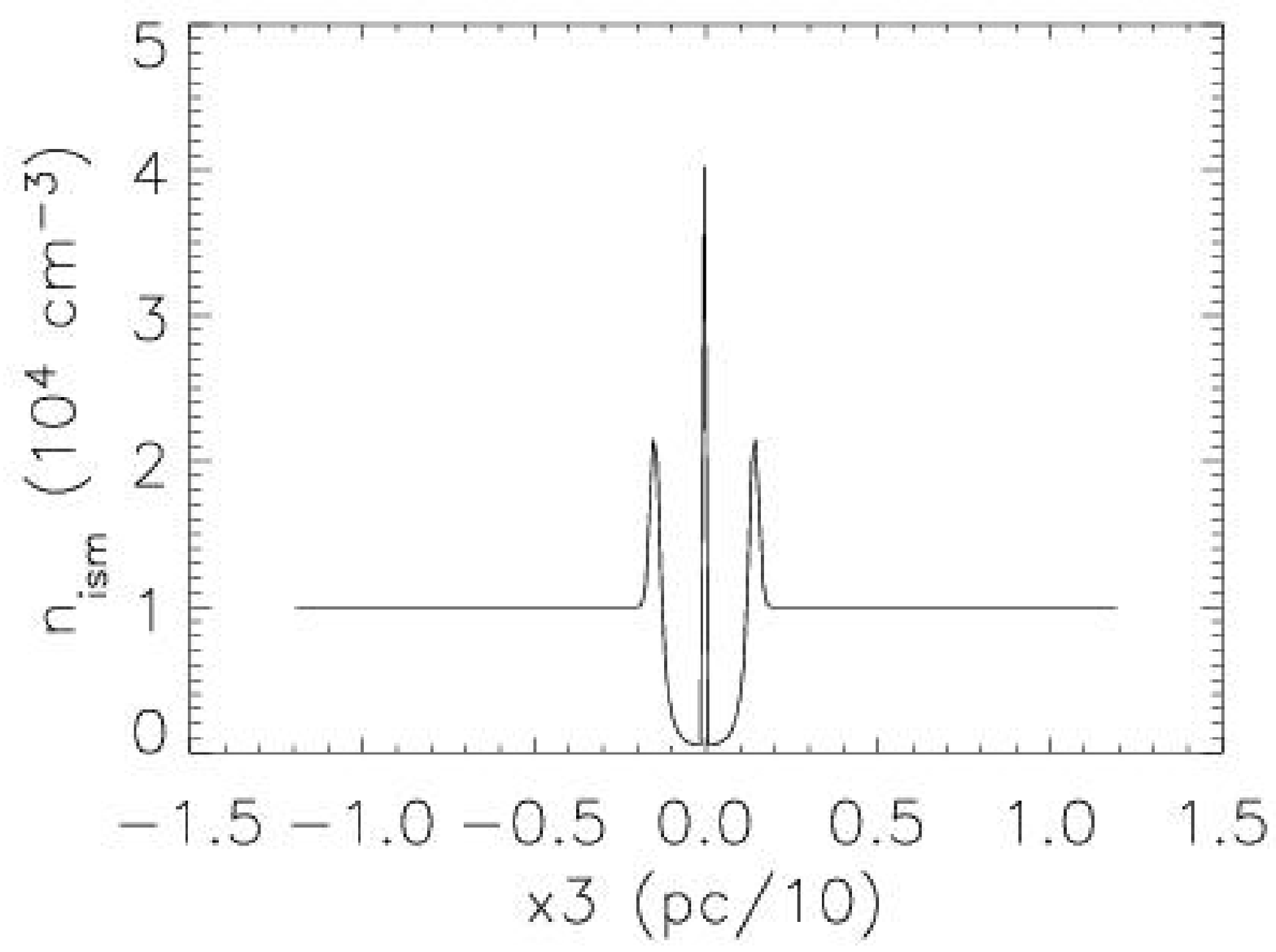}
\includegraphics{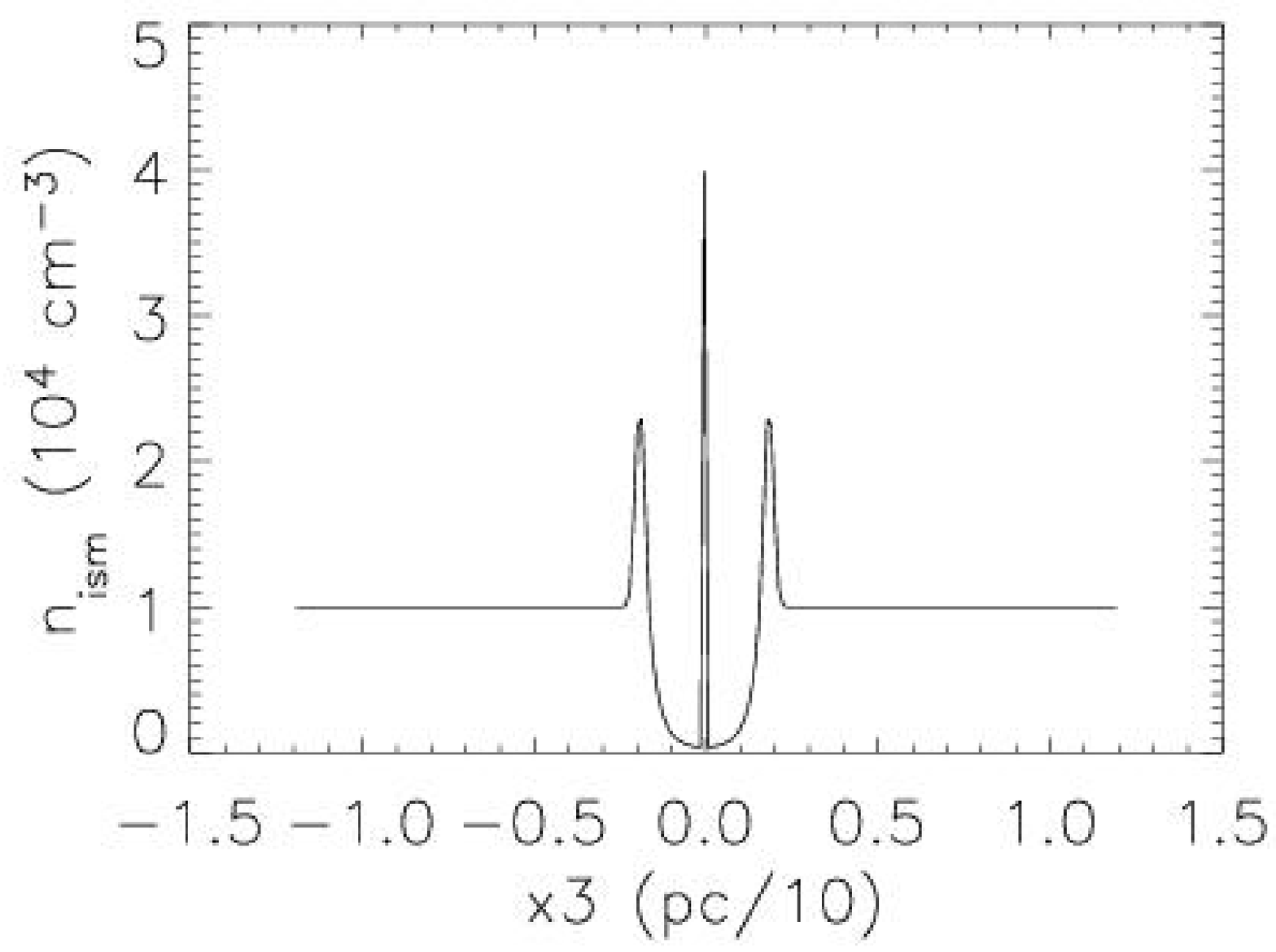}

\end{center}
\caption{Jet density is shown here as a function of time.  Each panel represents 800 (top left), 1200 (top right), 1600 (bottom left) and 2000 (bottom right) years into the $t_{\rm off}$ = $\infty$ case, at a distance of 0.2 pc from the source along the jet axis.  The density on the y axis is given in units of $10^4$ cm$^{-3}$, while the units on the x axis are given in units of 0.1 pc from  the jet axis.}
\label{fig:dens}
\end{figure}

\end{document}